\documentclass[11pt,3p,times]{elsarticle}

\usepackage{amsmath,amsfonts,amssymb}
\usepackage{amsthm}
\usepackage{bm}
\usepackage{graphicx,epstopdf}
\usepackage[position=top,labelfont=normalfont,textfont=normalfont,singlelinecheck=off,justification=raggedright]{subfig}
\usepackage{floatrow}
\usepackage[dvipsnames]{xcolor}
\usepackage{setspace}
\usepackage{enumerate,etaremune}
\usepackage{booktabs}
\usepackage{tabularx,multirow}
\usepackage{framed}
\usepackage{hhline}
\usepackage{url}
\usepackage[unicode,bookmarks=false]{hyperref}
\hypersetup{
  colorlinks,
  citecolor=Blue,linkcolor=Blue,urlcolor=Blue
}
\usepackage[toc,page]{appendix}
\usepackage[T1]{fontenc}
\usepackage{cancel}

\usepackage{lineno}

\usepackage[textsize=footnotesize,linecolor=red,backgroundcolor=red!25,bordercolor=red]
  {todonotes}

\usepackage{tikz}
\usetikzlibrary{shapes.geometric}
\usetikzlibrary{shapes,arrows}
\usetikzlibrary{plotmarks}

\usetikzlibrary{calc}

\usepackage{algorithm}
\usepackage{algorithmicx,algpseudocode}
\usepackage{cases}

\newcommand{\eg}{{\it e.g.}}
\newcommand{\ie}{{\it i.e.}}

\newcommand{\etal}{{\it et al.}}
\newcommand{\tensor}[1]{\bm{#1}}

\newcommand{\dd}{\mathrm{d}}

\newcommand{\rn}[1]{\uppercase\expandafter{\romannumeral #1\relax}}

\DeclareMathOperator{\diver}{\nabla\cdot}
\DeclareMathOperator{\symgrad}{\nabla^{s}}

\DeclareMathOperator{\trace}{tr}

\DeclareMathOperator*{\assembly}{\bm{\mathsf{A}}}
\newsavebox{\dotbox}

\theoremstyle{remark}
\newtheorem{remark}{Remark}

\newcommand{\revised}[1]{{\color{black} #1}}

\newcolumntype{L}[1]{>{\raggedright\let\newline\\arraybackslash\hspace{0pt}}m{#1}}
\newcolumntype{C}[1]{>{\centering\let\newline\\arraybackslash\hspace{0pt}}m{#1}}
\newcolumntype{R}[1]{>{\raggedleft\let\newline\\arraybackslash\hspace{0pt}}m{#1}}

\allowdisplaybreaks

\biboptions{sort&compress,square,comma,numbers}
\AtBeginDocument{\hypersetup{citecolor=MidnightBlue,linkcolor=MidnightBlue,urlcolor=MidnightBlue}}

\usepackage{etoolbox}
\makeatletter
\patchcmd{\ps@pprintTitle}
{Preprint submitted to}
{~}
{}{}
\makeatother
\journal{~}

\begin{document}

\begin{frontmatter}

\title{GeoWarp: An automatically differentiable and GPU-accelerated implicit MPM framework for geomechanics based on NVIDIA Warp}


\author[KAIST]{Yidong Zhao}
\author[UCLA]{Xuan Li}
\author[UCLA]{Chenfanfu Jiang}
\author[SNU-CEE,SNU-ICE]{Jinhyun Choo\corref{corr}}
\ead{jinhyun.choo@snu.ac.kr}

\cortext[corr]{Corresponding Author}

\address[KAIST]{Department of Civil and Environmental Engineering, KAIST, Daejeon, South Korea}
\address[UCLA]{Department of Mathematics, University of California, Los Angeles, United States}
\address[SNU-CEE]{Department of Civil and Environmental Engineering, Seoul National University, Seoul, South Korea}
\address[SNU-ICE]{Institute of Construction and Environmental Engineering, Seoul National University, Seoul, South Korea}


\begin{abstract}
The material point method (MPM), a hybrid Lagrangian--Eulerian particle method, is increasingly used to simulate large-deformation and history-dependent behavior of geomaterials. 
While explicit time integration dominates current MPM implementations due to its algorithmic simplicity, such schemes are unsuitable for quasi-static and long-term processes typical in geomechanics. 
Implicit MPM formulations are free of these limitations but remain less adopted, largely due to the difficulty of computing the Jacobian matrix required for Newton-type solvers, especially when consistent tangent operators should be derived for complex constitutive models. 
In this paper, we introduce GeoWarp---an implicit MPM framework for geomechanics built on NVIDIA Warp---that exploits GPU parallelism and reverse-mode automatic differentiation to compute Jacobians without manual derivation.
To enhance efficiency, we develop a sparse Jacobian construction algorithm that leverages the localized particle--grid interactions intrinsic to MPM. 
The framework is verified through forward and inverse examples in large-deformation elastoplasticity and coupled poromechanics. 
Results demonstrate that GeoWarp provides a robust, scalable, and extensible platform for differentiable implicit MPM simulation in computational geomechanics.
\end{abstract}


\end{frontmatter}


\section{Introduction}
The material point method (MPM)~\cite{sulsky1994particle} is a continuum particle method that combines Lagrangian and Eulerian descriptions.
It tracks the state of materials using a set of particles in a Lagrangian framework, while solving the governing equations on a background grid following an Eulerian approach.
This hybrid Lagrangian--Eulerian formulation allows MPM to simulate large-deformation kinematics of history-dependent materials without suffering from mesh distortion.
This capability makes MPM particularly well suited for modeling geomaterials (\eg~soils, rocks, and snow) that undergo large deformations.
As a result, it has been increasingly adopted in computational geomechanics to model the behavior of geomaterials and their interactions with various objects and processes (\eg~\cite{zabala2011progressive,yerro2015material,kularathna2021semi,dunatunga2015continuum,gaume2018dynamic,li2021three,seyedan2021solid,dunatunga2022modelling,jiang2022hybrid,jiang2023erosion,liang2023revealing,zhao2023coupled,yu2024semi,zhao2024mapped}).

Most existing MPM implementations employ explicit time integration schemes---particularly the explicit Euler method---due to their algorithmic simplicity, ease of implementation, and suitability for dynamic problems.
Accordingly, explicit MPM has been widely adopted in open-source codes such as
Uintah-MPM~\cite{germain2000uintah},
NairnMPM~\cite{nairn2015},
CB-Geo MPM~\cite{kumar2019scalable},
Karamelo~\cite{de2021karamelo},
fMPM-solver~\cite{wyser2020fast},
GeoTaich~\cite{shi2024geotaichi},
Anura3D~\cite{Anura3D2025},
MaterialPointSolver.jl~\cite{huo2025high}, and
Matter~\cite{blatny2025matter}.
Despite their advantages, explicit schemes are only conditionally stable and impose severe limitations on the time step size. 
Furthermore, explicit methods do not allow control over numerical error per time step.
Due to these reasons, explicit MPM is undesirable for quasi-static or long-term problems prevalent in geomechanics~\cite{lacroix2020life}.

Implicit time integration schemes, in contrast, are unconditionally stable and permit error control at each time step. 
These features make them more suitable for quasi-static or long-term processes, and implicit solvers have been widely employed in geomechanics (\eg~\cite{young2009liquefaction,borja2010continuum,choo2015stabilized,choo2016hydromechanical,choo2018cracking}). 
However, implementing implicit MPM is considerably more challenging, as it requires solving a nonlinear system of equations at every time step. 
This is typically achieved using Newton’s method, which in turn demands accurate derivation and implementation of the Jacobian matrix to ensure convergence.
Deriving the Jacobian matrix is particularly tedious and error-prone when the formulation involves complex constitutive models, as the consistent tangent operator---defined as the derivative of stress with respect to strain consistent with the time integration algorithm---should also be computed~\cite{simo1985consistent}. 
In the context of MPM, only a limited number of studies have pursued implicit formulations (\eg~\cite{guilkey2003implicit,beuth2011solution,wang2016development,charlton2017igimp,coombs2020ample,coombs2020lagrangian,zhao2020stabilized,chandra2024mixed}). 
Among these, AMPLE~\cite{coombs2020ample} provides an open-source implementation of implicit MPM with clear documentation and example applications. 
However, as noted by the authors, AMPLE is primarily intended for conceptual demonstration and is not designed for high-performance computing or large-scale simulation.
To date, no open-source, high-performance implementation of implicit MPM tailored specifically for computational geomechanics has been made available.

Automatic differentiation (AD) offers a promising pathway toward implementing implicit solvers and enabling differentiable simulation. By constructing a computational graph during the forward pass and applying the chain rule in reverse during the backward pass, AD allows for the exact and automatic computation of gradients. This capability has significantly impacted simulation-based fields, including computational mechanics, where AD is increasingly used to simplify the implementation of nonlinear solvers.
In the context of the finite element method (FEM), recent studies have demonstrated that AD can greatly reduce development effort by eliminating the need to manually derive and implement Jacobian matrices for complex, path-dependent constitutive models (\eg~\cite{vigliotti2021automatic,xue2023jax,dummer2024robust}). 
Beyond implementation efficiency, AD also enables gradient-based optimization workflows, including inverse analysis, control, and learning.
In the MPM community, several recent frameworks have successfully integrated AD with explicit MPM solvers. 
For example, ChainQueen~\cite{hu2019chainqueen}, DiffTaichi~\cite{hu2019difftaichi}, and PlasticineLab~\cite{huang2021plasticinelab} demonstrate differentiable simulations for soft robotics and control applications by embedding AD into GPU-accelerated MPM engines. 
However, these frameworks are primarily built for dynamic simulations using explicit time integration. 
The extension of AD to implicit MPM remains largely unexplored, despite its potential to simplify the implementation of Newton-type solvers and support inverse modeling in geomechanical problems.

Building on recent advances in AD, here we introduce GeoWarp---an open-source, high-performance framework for implicit MPM tailored to computational geomechanics built on NVIDIA Warp~\cite{warp2022}.
Warp is a Python-based simulation platform that supports AD and just-in-time compilation for efficient execution on both CPUs and GPUs. 
Its differentiable kernel-based programming model enables automatic and accurate gradient computation, while its Python interface facilitates rapid development. 
These features make Warp a suitable foundation for constructing differentiable solvers for large-scale mechanics problems.

At the core of GeoWarp is a fully implicit MPM solver integrated with AD. 
This integration eliminates the need for manually deriving and implementing Jacobian matrices, which are required for Newton-type solvers. 
In geomechanics, where constitutive models are often nonlinear and history-dependent, consistent tangent operators can be particularly difficult to derive and implement correctly. 
By recording the full computational graph of the simulation, GeoWarp enables automatic Jacobian computation, substantially lowering the barrier to implementing robust implicit MPM methods.

To address the computational cost typically associated with AD, we introduce a Jacobian construction strategy that leverages the sparsity pattern inherent in MPM. 
By exploiting the locality of particle--grid interactions, the method reduces reverse-mode evaluations to a small, fixed number of backward passes independent of problem size. 
This algorithm enables efficient large-scale simulation on modern GPUs, even for three-dimensional problems.
Beyond forward simulation, the automatic differentiability of GeoWarp facilitates inverse modeling tasks such as material parameter identification and integration with gradient-based optimization and learning-based methods. 

The framework is verified through forward and inverse examples in large-deformation elastoplasticity and coupled poromechanics. 
Results demonstrate that GeoWarp provides a robust, scalable, and extensible platform for differentiable implicit MPM simulations in computational geomechanics. 
The implementation is released as an open-source codebase to support reproducibility and future development: \url{https://github.com/choo-group/GeoWarp}.

\section{Material point method formulation}
\label{sec:mpm_formulation}
This section outlines the implicit MPM formulation adopted in this work. 
We begin by stating the initial--boundary value problem governing large-deformation mechanics, followed by its spatial discretization using the material point method. 
The resulting nonlinear system is then solved using a fully implicit scheme with Newton’s method. 
For brevity, we focus on describing the essential components of MPM, referring readers to comprehensive references for further details~\cite{jiang2016material,zhang2016material,fern2019material,nguyen2023material}.

\subsection{Problem statement}
Consider a deformable body occupying a domain $\Omega$ with boundary $\partial \Omega$. 
The boundary is partitioned into a Dirichlet portion $\partial_{\tensor{u}} \Omega$ where displacements are prescribed, and a Neumann portion $\partial_{\tensor{t}} \Omega$ where tractions are applied. 
These subsets satisfy the standard conditions $\partial_{\tensor{u}} \Omega \cap \partial_{\tensor{t}} \Omega = \emptyset$ and $\overline{\partial_{\tensor{u}} \Omega \cup \partial_{\tensor{t}} \Omega} = \partial \Omega$. 
The problem is defined over a time interval $\mathcal{T} := (0, T]$, with $T > 0$.

To accommodate large deformations, we adopt a finite deformation framework that distinguishes between the reference and current configurations. 
A material point is identified by its position vector in the reference configuration, $\tensor{X}$, and its position vector in the current configuration, $\tensor{x}$.
The displacement vector is then defined as $\tensor{u} := \tensor{x} - \tensor{X}$.
The deformation gradient is defined as
\begin{equation}
    \tensor{F} := \dfrac{\partial \tensor{x}}{\partial \tensor{X}} = \tensor{1} + \dfrac{\partial \tensor{u}}{\partial \tensor{X}},
\end{equation}
where $\tensor{1}$ denotes the second-order identity tensor.
The Jacobian,
\begin{equation}
    J := \det(\tensor{F}),
\end{equation}
represents the ratio of the current differential volume, $\dd v$, to the reference differential volume, $\dd V$.

In the current configuration, the momentum balance under quasi-static conditions is given by
\begin{equation}
    \diver \tensor{\sigma}(\tensor{F}) + \rho \tensor{g} = \tensor{0} \quad\text{in}\quad \Omega \times \mathcal{T},
\end{equation}
where $\tensor{\sigma}$ is the Cauchy stress tensor, $\diver$ is the divergence operator evaluated in the current configuration, $\rho$ is the current mass density, and $\tensor{g}$ is the gravitational acceleration vector.
The Dirichlet and Neumann boundary conditions are prescribed as
\begin{align}
    \tensor{u} = \hat{\tensor{u}} \quad&\text{on}\quad \partial \Omega_{\tensor{u}} \times \mathcal{T},\\
    \tensor{n} \cdot \tensor{\sigma} = \hat{\tensor{t}} \quad&\text{on}\quad \partial \Omega_{\tensor{t}} \times \mathcal{T},
\end{align}
with $\hat{\tensor{u}}$ and $\hat{\tensor{t}}$ denoting the prescribed boundary displacement and traction, respectively. 
The initial condition is given by
\begin{equation}
    \tensor{u} = \tensor{u}_0 \quad\text{at}\quad t = 0,
\end{equation}
where $\tensor{u}_0$ denotes the initial displacement field. 
To close the formulation, a constitutive law relating the Cauchy stress tensor $\tensor{\sigma}$ to the deformation gradient $\tensor{F}$ should be specified. 
In this study, we consider several constitutive models commonly used in geomechanics. 
For brevity, we omit their detailed formulations and refer the reader to standard references on constitutive models (\eg~\cite{de2011computational,borja2013plasticity}).

\subsection{Implicit MPM discretization}
For the MPM discretization of the problem, we represent the domain as a collection of particles (material points) and introduce a background computational grid that interacts with the particles. 
The particles carry state variables such as stress and volume, following a Lagrangian description. 
In contrast, the governing equations are discretized and solved on the background grid, which follows an Eulerian description. 
To couple these two descriptions, field quantities are transferred between the particles and grid nodes.
Let us use the subscripts $(\cdot)p$ and $(\cdot)i$ to denote quantities associated with particle $p$ and grid node $i$, respectively. 
The mapping of an arbitrary variable $f$ from particles to nodes can be written as
\begin{equation}
    f_{i} = \sum_{p=1}^{n_{p}}w_{i,p}m_{p}f_{p}/M_{i}\,, \quad
    M_{i} := \sum_{p=1}^{n_{p}} w_{i,p}m_{p}\,.
    \label{eq:map-from-particles-to-nodes}
\end{equation}
where $n_p$ is the number of material points influencing node $i$, $m_p$ is the mass of particle $p$, and $w_{i,p}$ is the weighting function associated with node $i$ evaluated at the position of particle $p$. These weighting functions serve the same role as shape functions in the finite element method.

For the MPM weighting functions, the original MPM formulation~\cite{sulsky1994particle} employs linear finite element shape functions; however, such functions are prone to interpolation errors when particles cross cell boundaries. 
To alleviate this issue, several enhanced interpolation schemes have been developed, including the Generalized Interpolation Material Point (GIMP)~\cite{bardenhagen2004generalized} method, Convected Particle Domain Interpolation (CPDI)~\cite{sadeghirad2011convected,sadeghirad2013second}, B-spline-based algorithms~\cite{steffen2008examination,steffen2008analysis}, and Moving Least Squares (MLS) interpolation~\cite{hu2018moving}.
In this work, we adopt the contiguous GIMP (cpGIMP) scheme~\cite{wallstedt2008evaluation}, which has been widely used in conjunction with implicit MPM formulations~\cite{charlton2017igimp,coombs2020ample,coombs2020lagrangian}. 
The one-dimensional weighting function and its gradient follow the formulation presented by Coombs and Augarde~\cite{coombs2020ample}. 
In multiple dimensions, the weighting functions and their gradients are constructed as tensor products of the corresponding one-dimensional components along each spatial direction.

\begin{figure}[h!]
    \centering
    \includegraphics[width=1\textwidth]{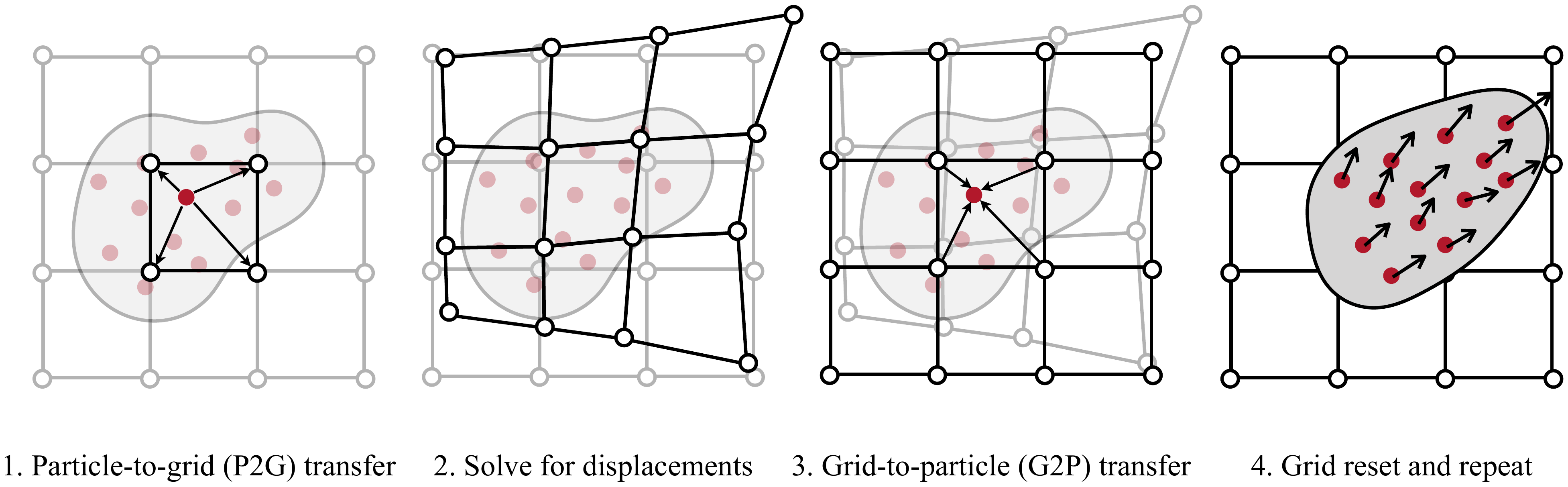}
    \caption{MPM update procedure.}
    \label{fig:mpm_procedure}
\end{figure}

At each time step, the MPM update follows a four-stage procedure, as illustrated in Fig.~\ref{fig:mpm_procedure} and described below:
\begin{enumerate}
    \item Particle-to-grid (P2G) transfer: Particle quantities are mapped to the background grid using the selected weighting functions.
    \item Nodal update: The governing equations are solved on the grid to compute nodal displacements (or velocities, depending on the formulation).
    \item Grid-to-particle (G2P) transfer: Updated nodal values are interpolated back to the particles to recover particle-level kinematics.
    \item Particle update: Particle positions and state variables are updated accordingly. The background grid is then reset in preparation for the next time step.
\end{enumerate}
Once the above procedure is completed, the simulation advances to the next time step, repeating the four-stage process until the end of the analysis.

In the second step, the governing equation is solved in its variational form. 
Applying the standard procedure to the momentum balance gives
\begin{equation}
    -\int_{\Omega} \symgrad \tensor{\eta} : \tensor{\sigma} \, \dd v + \int_{\Omega} \tensor{\eta} \cdot \rho \tensor{g} \, \dd v + \int_{\partial \Omega_{\tensor{t}}} \tensor{\eta} \cdot \hat{\tensor{t}} \, \dd a = 0,
    \label{eq:variational_form}
\end{equation}
where $\tensor{\eta}$ denotes the variation of the displacement field, $\symgrad$ denotes the symmetric gradient operator evaluated in the current configuration, and $\dd a$ denotes the differential surface area in the current configuration.
The variational equation can be discretized using the Galerkin method. 
The resulting discrete form is evaluated as
\begin{equation}
    \assembly_{p} \left(\tensor{f}_{{\rm int},p} (\tensor{u}_p)\right)
    = \assembly_{p} \left(\tensor{f}_{{\rm ext},p}\right),
\end{equation}
where $\assembly_{p}$ denotes the global assembly over all particles, and $\tensor{f}_{{\rm int},p}$ and $\tensor{f}_{{\rm ext},p}$ are the internal and external force vectors, respectively, defined as
\begin{align}
    \tensor{f}_{{\rm int},p} &:= \left( \symgrad\tensor{w}_{p} \right)^\top : \tensor{\sigma}_p V_p, \\
    \tensor{f}_{{\rm ext},p} &:= \tensor{w}_p^{\top} \rho_p \tensor{g} V_p + \tensor{w}_p^{\top} \hat{\tensor{t}}_p V_p,
\end{align}
with $\tensor{w}_p$ and $\symgrad \tensor{w}_p$ denoting the weighting functions and their symmetric gradients evaluated at grid nodes influenced by particle $p$, and $V_p$ denoting the particle volume.

In this work, we solve the discrete governing equation using the implicit (backward) Euler method.
Let superscripts $(\cdot)^n$ and $(\cdot)^{n+1}$ denote quantities at time steps $n$ and $n+1$, respectively. 
Given all relevant quantities at step~$n$, our goal is to compute the displacement field $\tensor{u}^{n+1}$ at step~$n+1$. 
Since the governing equation is generally nonlinear in $\tensor{u}^{n+1}$, we make use of Newton’s method to solve the residual equation
\begin{equation}
    \tensor{r}(\tensor{u}^{n+1}) = \assembly_{p}  \left(\tensor{f}_{{\rm int},p}(\tensor{u}_p^{n+1}) - \tensor{f}_{{\rm ext},p}^{n+1}\right) \rightarrow \tensor{0},
\end{equation} 
where $\tensor{r}(\tensor{u}^{n+1})$ is the global residual vector assembled from particle contributions.
At each iteration $k$, we compute the update by solving the linear system
\begin{equation}
    -\tensor{J}^{n+1,k} \Delta \tensor{u}^{n+1, \, k} = \tensor{r}(\tensor{u}^{n+1,k}),
\end{equation}
where the Jacobian matrix is defined as
\begin{equation}
    \tensor{J}^{n+1,k} := \dfrac{\partial\tensor{r}(\tensor{u}^{n+1,k})}{\partial \tensor{u}^{n+1,k}}.
\end{equation}
The iterations continue until the relative residual norm satisfies the prescribed convergence criterion:
\begin{equation}
    \dfrac{\Vert \tensor{r}(\tensor{u}^{n+1,k})\Vert}{\Vert \tensor{r}(\tensor{u}^{n+1,0})\Vert}\leq \text{tol},
\end{equation}
where \text{tol} denotes the tolerance. 
Once convergence is achieved, the displacement field is updated using the increment $\Delta \tensor{u}^{n+1,k}$.

The convergence behavior of Newton’s method is highly sensitive to the accuracy of the Jacobian matrix. 
In computational mechanics, the primary difficulty in evaluating the Jacobian lies in computing the consistent tangent operator, defined as the derivative of the incremental stress with respect to the incremental strain (or an equivalent measure)~\cite{simo1985consistent}. 
For relatively simple constitutive models, this operator can be derived analytically. 
However, for complex models commonly used in geomechanics, the derivation and implementation of consistent tangents is often tedious, error-prone, and model-specific. 
To overcome this limitation, we leverage AD to compute Jacobian matrices in a manner that is accurate, general, and fully automated. 
The details of our AD-based approach are presented in the following section.
\medskip

\remark{In addition to the single-phase mechanics formulation described above, we consider a coupled poromechanics formulation based on the $\tensor{u}$--$p$ formulation, which is widely used in geomechanics to model fluid flow and solid deformation in saturated porous media (\eg~\cite{white2008stabilized,choo2018enriched,choo2018large,choo2019stabilized}). 
This formulation introduces the pore pressure $p$ as an additional primary variable, augmenting the solid momentum balance with a fluid mass conservation equation. 
The resulting system captures two-way coupling between fluid diffusion and solid deformation.
The $\tensor{u}$--$p$ formulation is also implemented in GeoWarp based on the MPM formulation described in Zhao and Choo~\cite{zhao2020stabilized}. 
The reader is referred to that work for full details of the governing equations and numerical implementation.}
\medskip

\remark{In our current implementation, each Newton iteration solves a linear system using matrix-based methods that rely on explicit Jacobian assembly. 
For single-phase problems, we use a preconditioned Algebraic Multigrid iterative solver~\cite{pyamg2023}, while for coupled poromechanics, a sparse direct solver is employed~\cite{pypardiso}.
To further improve scalability and memory efficiency, future work may explore block-preconditioned Krylov solvers tailored for poromechanical systems (\eg~\cite{white2011block,white2016block}), as well as matrix-free Krylov methods that compute Jacobian–vector products on-the-fly via forward-mode AD.
\revised{The modified Newton's method that reuses previously assembled Jacobians for multiple iterations can also be explored.}}

\section{Automatic differentiation for Jacobian construction}
\label{sec:automatic_differentiation}

This section outlines the strategies employed in the GeoWarp framework to compute the Jacobian matrix automatically---a critical component for efficient implicit MPM computation and differentiable simulations. 
Leveraging Warp's reverse-mode AD, GeoWarp computes Jacobian matrices without manual derivation, which is particularly advantageous when complex constitutive models are used. As with other modern AD libraries (\eg~\cite{Abadi_TensorFlow_Large-scale_machine_2015,ansel2024pytorch,jax2025github}), the Jacobian is evaluated through a sequence of Jacobian-vector products of the form $\tensor{J}^{\top} \tensor{e}$, where $\tensor{e}$ is a prescribed seed vector.
This procedure consists of two stages:
\begin{enumerate}
	\item Forward pass: The residual $\tensor{r}(\tensor{u})$ is computed while Warp constructs a computational graph of the associated operations.
	\item Backward pass: Reverse-mode AD is applied to this graph, starting from a seed vector $\tensor{e}$, to obtain the product $\tensor{J}^\top \tensor{e}$.
\end{enumerate}
In what follows, we first introduce a baseline implementation, referred to as \emph{dense differentiation}, in which each row of the Jacobian is computed via a separate AD backward pass. While general, this algorithm is computationally expensive for large systems. To address this, we develop a novel algorithm termed \emph{sparse differentiation}, which exploits the localized nature of particle--grid interactions intrinsic to MPM. This method significantly reduces the number of required backward passes---from one per degree of freedom to a small, fixed number---thereby improving scalability and efficiency for large-scale simulations.

\subsection{Dense differentiation for Jacobian construction}
A direct, albeit computationally expensive, strategy for constructing the Jacobian matrix involves looping over all degrees of freedom. 
For each degree of freedom $i$, a seed vector $\tensor{e}$ is defined such that $\tensor{e}[i] = 1$ and all other components are zero. 
Applying reverse-mode automatic differentiation (AD) with this seed vector yields the $i$-th row of the Jacobian matrix $\tensor{J}$.
The procedure is summarized in Algorithm~\ref{algorithm:dense_differentiation}.
\begin{algorithm}[h!]
    \setstretch{1.15}
    \caption{Dense differentiation}
    \label{algorithm:dense_differentiation}
    \begin{algorithmic}[1]
        \State Initialize an empty Jacobian matrix \(\tensor{J} \leftarrow []\)
        \For{each degree of freedom $i$}
            \State Initialize seed vector \(\tensor{e} \leftarrow \tensor{0}\)
            \State Set the $i$-th component of $\tensor{e}$ to 1: \(\tensor{e}[i] = 1\)
            \State Compute the $i$-th row of the Jacobian \(\tensor{J}_{i,\,:} = \texttt{warp.backward}(\tensor{r}, \tensor{e})\) ($\tensor{r}$: the residual vector)
        \State Insert the Jacobian row \(\tensor{J}_{i,\,:}\) into \(\tensor{J}\)
        \EndFor
    \end{algorithmic}
\end{algorithm}

While straightforward to implement, this algorithm scales poorly for large-scale problems, as it requires one reverse-mode pass per degree of freedom. 
To address this inefficiency, we next present a sparse differentiation strategy that exploits the localized particle--grid interactions inherent to MPM to reduce the number of AD passes required for Jacobian construction.

\subsection{Sparse differentiation for Jacobian construction}
To address the prohibitive cost associated with dense differentiation, we develop an efficient differentiation algorithm that leverages the inherent sparsity of the MPM formulation.
In MPM, the global Jacobian matrix is naturally sparse due to the localized nature of particle--grid interactions. 
Each particle interacts with only a limited subset of grid nodes defined by its shape function support, and conversely, each grid node receives contributions from nearby particles only. 
This spatial locality yields a sparsity pattern in which each degree of freedom is coupled to a small, localized neighborhood of other degrees of freedom.

\begin{figure}[h!]
    \centering
    \subfloat[Particle interacting with Nodes 1, 2, and 3]{\includegraphics[width=0.5\textwidth]{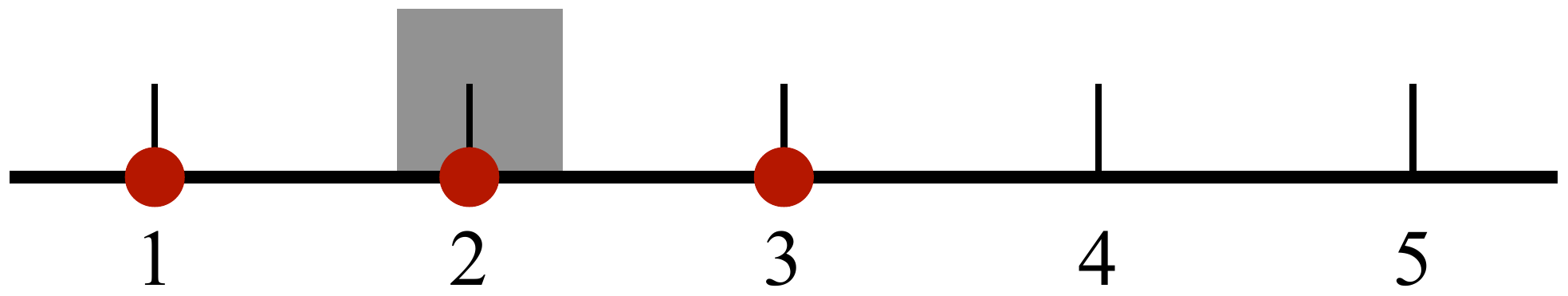}}\\
    \subfloat[Particle interacting with Nodes 2, 3, and 4]{\includegraphics[width=0.5\textwidth]{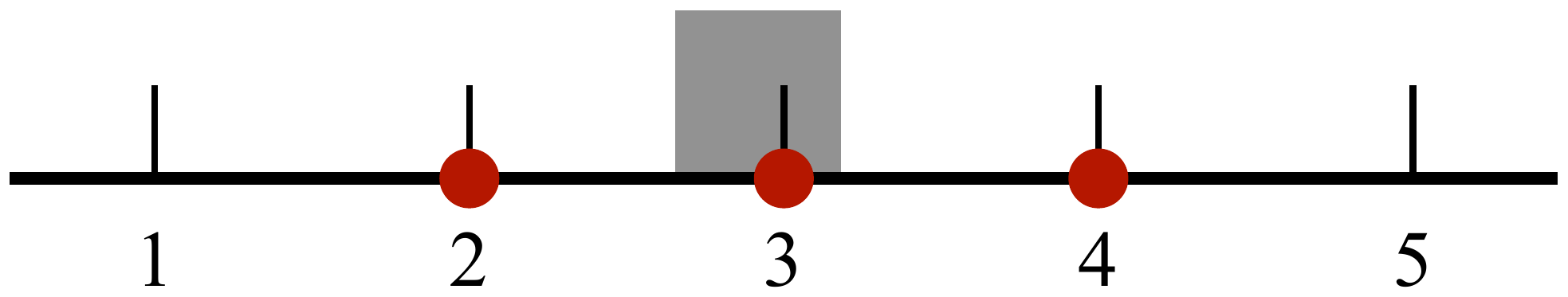}}\\
    \subfloat[Particle interacting with Nodes 3, 4, and 5]{\includegraphics[width=0.5\textwidth]{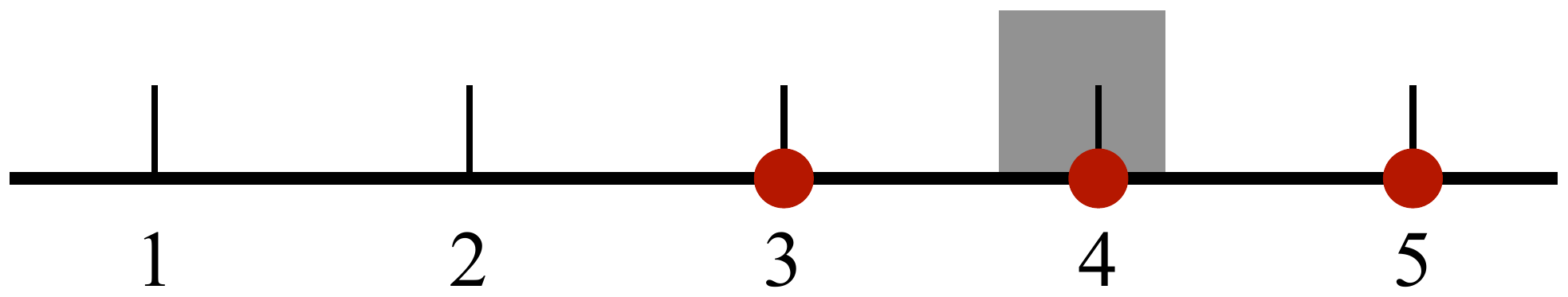}}
    \caption{Illustration of the associated grid nodes for a particle in the 1D GIMP formulation. The particle (shown as a gray rectangle) interacts with at most three neighboring grid nodes (marked as red dots) through its shape function support.}
    \label{fig:gimp_influence_domain}
\end{figure}

\begin{figure}[h!]
    \centering
    \includegraphics[width=0.5\textwidth]{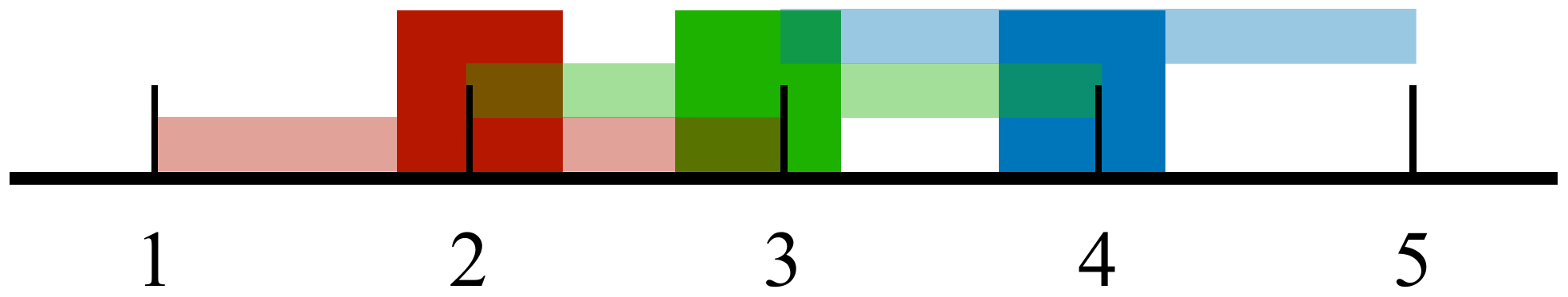}
    \caption{\revised{Overlapping influence domains of neighboring particles around Node 3. Rectangles represent three neighboring particles, and their shaded regions represent the respective influence zones.}}
    \label{fig:overlapping_influence_domains}
\end{figure}

To illustrate this structure, we consider the 1D GIMP formulation. 
In 1D GIMP, each particle is associated with a compact influence domain, and only the grid nodes within this domain participate in its transfer operations. 
For typical particle sizes and grid spacings, this domain spans a small number of nodes relative to the global domain. 
As shown in Fig.~\ref{fig:gimp_influence_domain}, a single particle (depicted as a gray rectangle) interacts with at most three grid nodes (shown as red dots). 
This localized coupling implies that each particle affects only a small portion of the global residual vector and contributes to only a few entries in the global Jacobian matrix.
As a result, the assembled Jacobian exhibits a block-sparse structure, with nonzero entries confined to submatrices corresponding to the particle's neighboring grid nodes. 
\revised{To further clarify this block-sparse structure, consider grid Node 3 in Fig.~\ref{fig:overlapping_influence_domains}. 
The red, green, and blue rectangles represent three neighboring particles around Node 3.
Their shaded regions denote the respective influence zones.
Note that although each particle interacts directly with at most three grid nodes, the overlap of their influence domains causes additional coupling among nearby nodes.
As a result, the largest nonzero submatrix of Node 3 spans Nodes 1 through 5 in the global Jacobian.
Extending this reasoning to higher dimensions, the overlap in both spatial directions produces a maximum nonzero block of $5 \times 5$ in 2D and $5 \times 5 \times 5$ in 3D.}

\begin{figure}[h!]
    \centering
    \subfloat[Dense differentiation]{\includegraphics[width=0.4\textwidth]{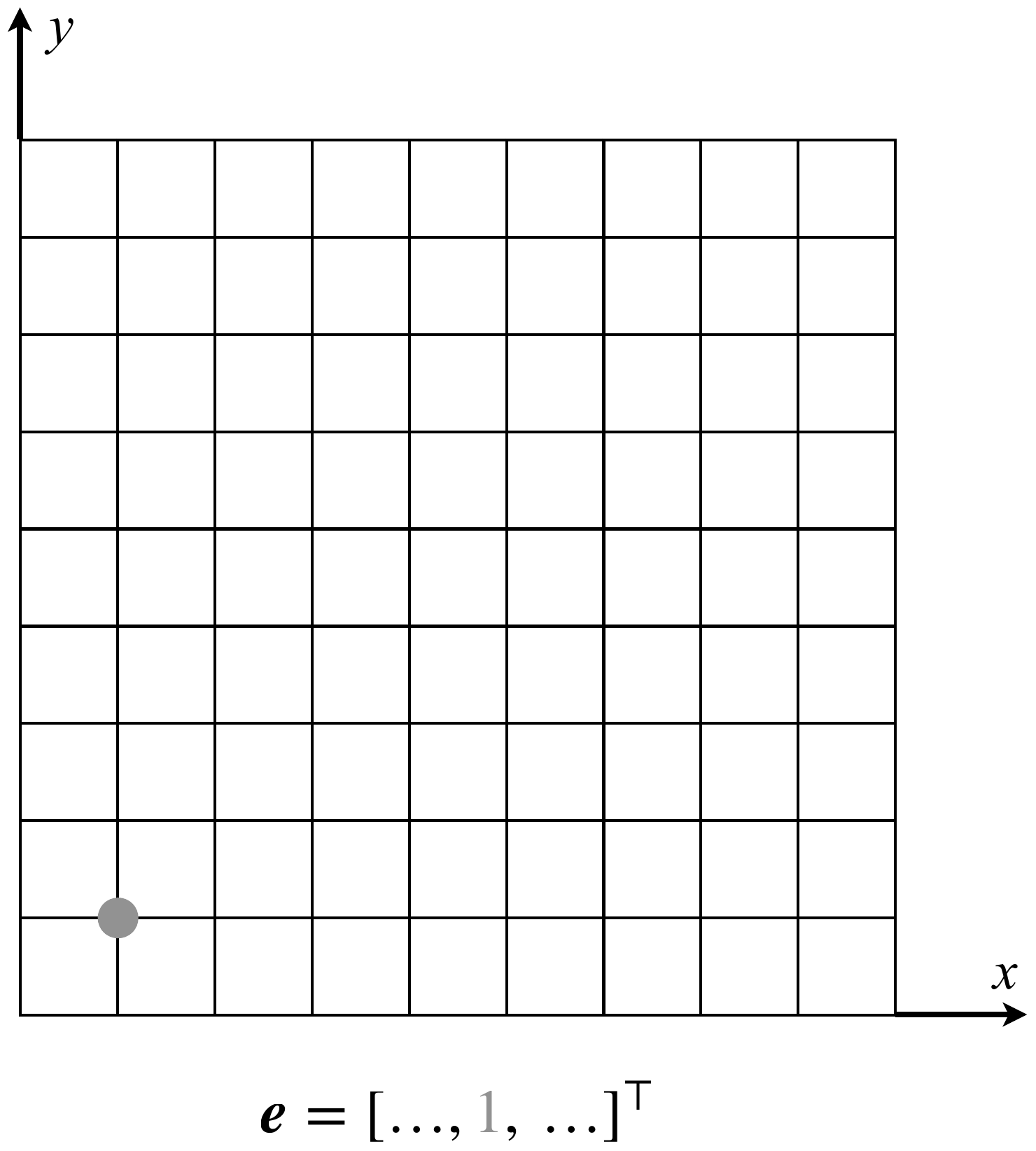}\label{fig:jacobian_construction_two_algorithms_dense}}\hspace{3em}
    \subfloat[\revised{Sparse differentiation}]{\includegraphics[width=0.4\textwidth]{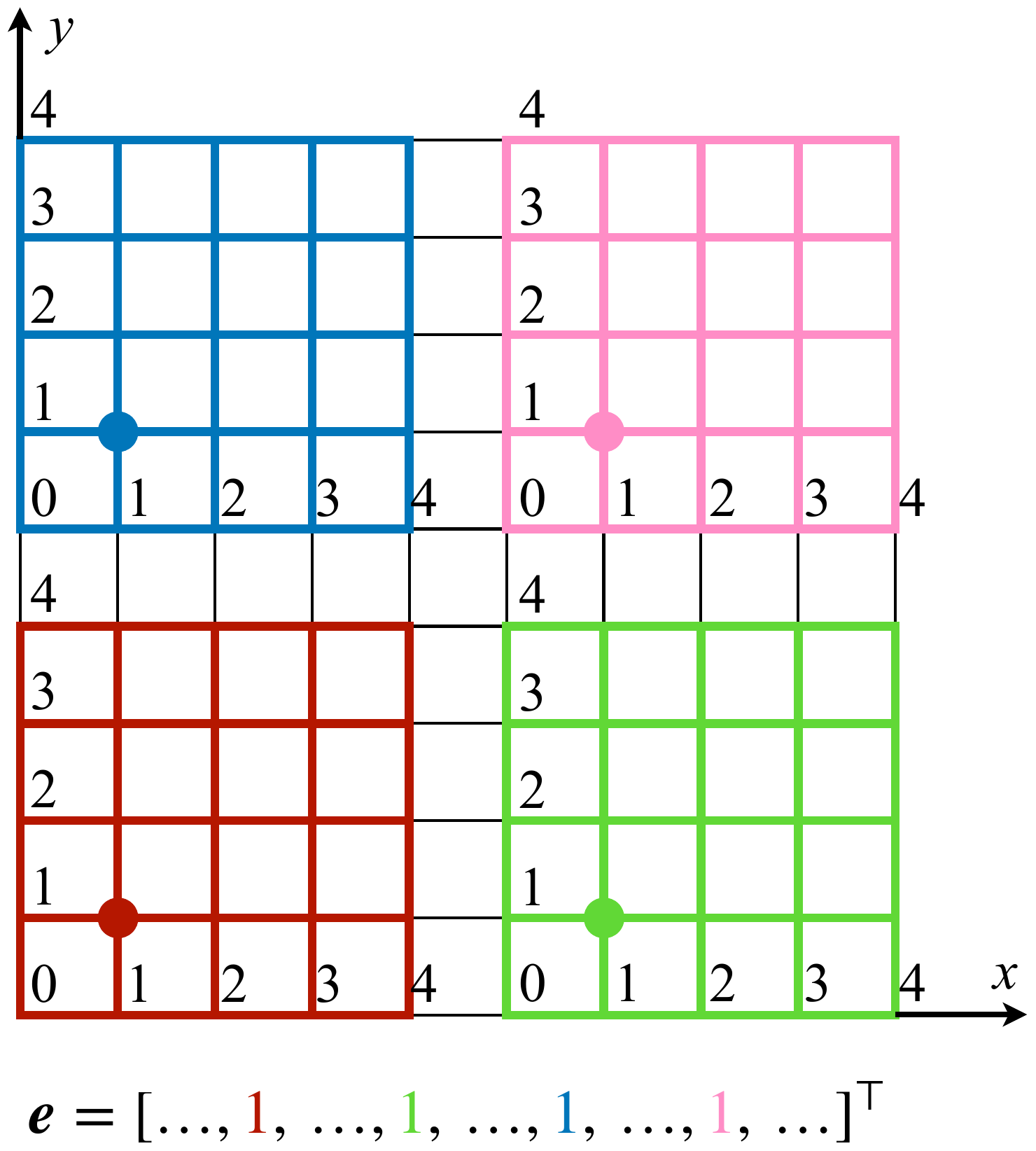}\label{fig:jacobian_construction_two_algorithms_sparse}}
    \caption{Two algorithms for Jacobian construction in implicit MPM. \revised{Solid dots indicate the active degrees of freedom.} (a) Dense differentiation requires one reverse-mode AD pass per degree of freedom, resulting in high computational cost. (b) Sparse differentiation exploits the locality of particle--grid interactions by partitioning the background grid into independent blocks, enabling multiple Jacobian rows to be computed simultaneously in a single backward pass. \revised{Numbers indicate the corresponding \texttt{local\_x} and \texttt{local\_y} indices.}}
    \label{fig:jacobian_construction_two_algorithms}
\end{figure}

This sparsity pattern enables a significantly more efficient algorithm for Jacobian construction. 
Rather than requiring one reverse-mode AD pass per degree of freedom---as in the dense differentiation---we compute a small, fixed number of Jacobian--vector products. 
In two dimensions, for example, the computational grid can be partitioned into a collection of non-overlapping $5 \times 5$ blocks, as illustrated in Fig.~\ref{fig:jacobian_construction_two_algorithms} by distinct color groupings. 
Each block corresponds to a spatially localized region with minimal interaction with adjacent blocks.

Exploiting this independence, we construct a seed vector $\tensor{e}$ with a single nonzero entry within each block, allowing the Jacobian rows associated with all active entries to be computed in parallel using a single backward pass. 
Because each seed entry is localized to a non-overlapping block, the resulting rows do not interfere with one another and can be extracted independently. 
This block-wise seeding strategy drastically reduces the number of backward passes required—down from one per degree of freedom to a small number determined by the block decomposition—and thereby accelerates the AD workflow while preserving exactness.

By iterating over all degrees of freedom using structured seed vectors, the full sparse Jacobian can be assembled with a fixed number of reverse-mode AD passes, determined by the size of the independent blocks.
\revised{The block size $b$ can be obtained through $b = 2 w - 1$, where $w$ is the support of the shape function (\ie~the number of grid nodes influenced by a particle along one spatial direction).} 
In this study, the GIMP scheme \revised{($w = 3$)} results in \revised{$b = 5$}, \revised{which corresponds to} 25 backward passes in 2D and 125 in 3D. 
This sparse differentiation strategy is not limited to GIMP and extends to other shape functions, with the number of AD passes varying according to the block size.
\revised{As summarized in Table~\ref{tab:block_size}, linear shape functions ($w = 2$) result in $b = 3$, while quadratic and cubic B-splines ($w = 3$ and $w = 4$) lead to $b = 5$ and $b = 7$, respectively. The total number of AD passes thus depends solely on the shape function support rather than the total number of unknowns in the global system. This dependence effectively decouples differentiation cost from global system size, which is critical for achieving scalability in large-scale three-dimensional simulations.}
This strategy enables substantial computational savings by leveraging the locality of particle--grid interactions.
Furthermore, the use of non-overlapping seed patterns facilitates fine-grained parallelism, allowing the differentiation process to be executed efficiently on modern hardware accelerators. 
The result is a substantial reduction in computational cost without compromising accuracy.
\begin{table}[h!]
    \centering
    \begin{tabular}{ccc}
    \hline
    Shape function & Shape function support $w$ & Block size $b = 2w - 1$  \\
    \hline
    Linear & 2 & 3 \\
    GIMP & 3 & 5 \\
    Quadratic B-splines & 3 & 5 \\
    Cubic B-splines & 4 & 7 \\
    \hline
    \end{tabular}
    \caption{\revised{Commonly used shape functions and corresponding block sizes.}}
    \label{tab:block_size}
\end{table}

The proposed sparse differentiation is summarized in Algorithm~\ref{algorithm:sparse_differentiation}, which presents the general procedure for three-dimensional problems. 
The two-dimensional case is recovered by omitting components in the $z$-direction.
\revised{The local indices \texttt{local\_x}, \texttt{local\_y}, and \texttt{local\_z} are introduced to systematically traverse the degrees of freedom within each independent block.
For example, the active degrees of freedom in Fig.~\ref{fig:jacobian_construction_two_algorithms_sparse} correspond to $\texttt{local\_x} = 1$ and $\texttt{local\_y} = 1$.}
\begin{algorithm}[h!]
    \setstretch{1.15}
    \caption{Sparse differentiation}
    \label{algorithm:sparse_differentiation}
    \begin{algorithmic}[1]
        \State Determine the block size \(b\) based on the chosen shape function \revised{shown in Table~\ref{tab:block_size}}
        \State Initialize an empty Jacobian matrix: \(\tensor{J} \gets []\)
        \For{each \texttt{local\_x} \(= 1, \ldots, b\)}
        \For{each \texttt{local\_y} \(= 1, \ldots, b\)}
        \For{each \texttt{local\_z} \(= 1, \ldots, b\)}
            \State Initialize seed vector: \(\tensor{e} \gets \tensor{0}\)
            \State Initialize active index set: \(\mathcal{I} \gets []\)
            \For{each block}
                \State Identify the degree of freedom corresponding to (\texttt{local\_x}, \texttt{local\_y}, \texttt{local\_z}) in the block
                \State Append the identified index to \(\mathcal{I}\)
            \EndFor
            \State Set \(\tensor{e}[\mathcal{I}] \gets 1\)
            \State Compute Jacobian rows: \(\tensor{J}_{\mathcal{I},\,:} \gets \texttt{warp.backward}(\tensor{r}, \tensor{e})\) (\(\tensor{r}\): the residual vector)
            \State Insert \(\tensor{J}_{\mathcal{I},\,:}\) into \(\tensor{J}\)
        \EndFor
        \EndFor
        \EndFor
    \end{algorithmic}
\end{algorithm}

\revised{Lastly, Warp provides a convenient finite-difference utility, $\texttt{warp.autograd.jacobian\_{fd}}$, that aids in debugging gradient computations. This function computes numerical approximations of derivatives, which can be directly compared with those obtained via automatic differentiation. Such tools are particularly useful for identifying issues related to non-symmetric Jacobian matrices.}

\section{Numerical examples}
\label{sec:numerical_examples}
In this section, we verify and demonstrate the capabilities of the GeoWarp framework through five numerical examples of increasing complexity.
The first example verifies the automatic computation of the consistent tangent operator via stress-point simulations of triaxial compression tests on sand using a critical-state plasticity model.
The second example verifies the MPM formulation and its implementation against analytical solutions for vertical bar compaction under self-weight.
The third example demonstrates the efficiency of the proposed sparse automatic differentiation algorithm for Jacobian assembly by simulating the deformation of a cantilever beam under a point load.
The fourth example verifies the implementation of the coupled $\tensor{u}$--$p$ formulation through one-dimensional consolidation simulations under both small and large deformations.
The fifth and final example demonstrates inverse analysis via differentiable simulation, identifying material parameters from the indentation response of a rigid footing into a saturated porous medium.
Together, these examples verify the framework’s accuracy, computational efficiency, and differentiability across both forward and inverse simulation contexts.

\subsection{Stress-point triaxial compression}
Stress-point simulations play a central role in geomechanics for the development, calibration, and validation of constitutive models.
In such simulations, Newton’s method is commonly employed to impose stress-controlled loading paths, for which a consistent tangent operator is essential to ensure robust and efficient convergence (see, \eg~\cite{borja2016cam,choo2018mohr}).
However, deriving the consistent tangent operator manually can be labor-intensive and error-prone, particularly for constitutive models exhibiting strong nonlinearity or complex state-dependent behavior.
In this example, we demonstrate GeoWarp’s capability to automatically compute the consistent tangent operator.

As an example of a complex constitutive model, we consider the Nor-Sand model~\cite{jefferies1993nor,jefferies2005norsand}, a widely used critical-state plasticity model for sands.
Since the original Nor-Sand formulation is rigid-plastic, we adopt the extended version by Borja and Andrade~\cite{borja2006critical}, which augments the model with pressure-dependent hyperelasticity and other enhancements.
The yield function is defined as
\begin{equation}
    f(p,q) = q + \eta p \leq 0,
\end{equation}
where $p = (1/3)\trace(\tensor{\sigma})$ is the mean normal stress, and $q = (\sqrt{3/2}) \Vert \tensor{\sigma} - p \tensor{1} \Vert$ is the deviatoric stress. 
(Unless otherwise noted, all normal stresses are understood to represent effective stresses, as appropriate for fluid-infiltrated geomaterials.)
The stress ratio $\eta$, which governs the size and shape of the yield surface, evolves according to
\begin{equation}
    \eta = \begin{cases}
    M \left[1 + \ln(p_i/p)\right] \; \; &\text{if } N = 0,\\
    (M/N) \left[1 - (1-N) (p/p_i)^{N/(1-N)}\right] \quad &\text{if } N > 0,
    \end{cases}
\end{equation}
where $M$ is the slope of the critical state line (CSL), $N$ is a curvature parameter, and $p_i$ is the image pressure, which represents the mean effective stress corresponding to the current specific volume on the CSL. 
The image pressure represents the mean effective stress corresponding to the current specific volume on the CSL and is implicitly determined through the critical state relation.

\begin{figure}[h!]
  \centering
  \includegraphics[width=0.24\textwidth]{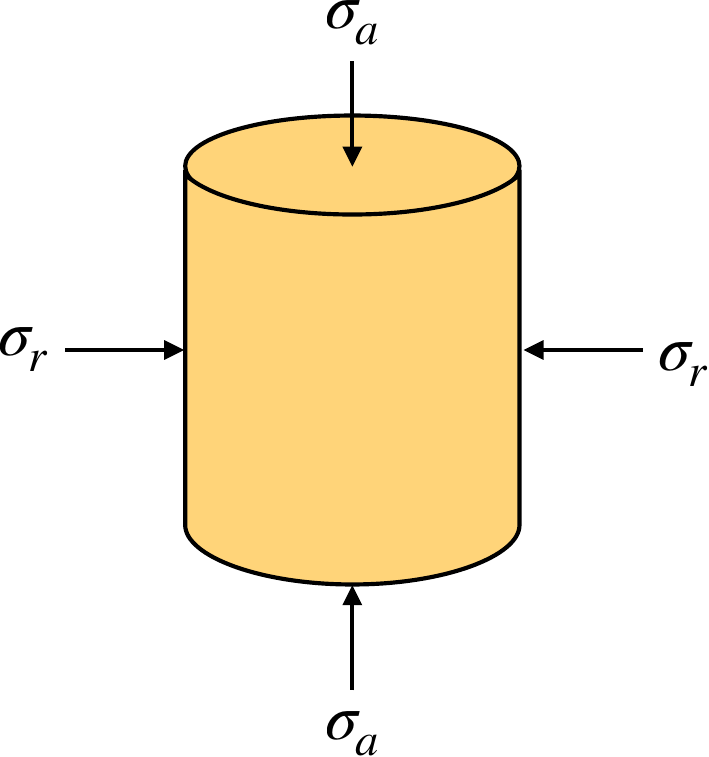}
  \caption{\revised{Stress-point triaxial compression: problem setup. The axial stress ($\sigma_a$) increases with loading, while the radial stress ($\sigma_r$) is kept constant.}}
  \label{fig:triaxial_setup}
\end{figure}

In this example, we adopt material parameters calibrated to drained triaxial compression tests on loose and dense Brasted sands~\cite{cornforth1964some}, as reported in Andrade and Ellison~\cite{andrade2008evaluation}.
The critical state parameters are set to $M = 1.27$ and $N = 0.4$, and the initial image pressure is specified as $-332.30$ kPa for the loose sand and $-534.47$ kPa for the dense sand.
The elastic compressibility is defined by $\tilde{\lambda} = 0.02$, with a reference specific volume $v_{c0} = 1.8911$. 
The initial specific volume is set to $v_0 = 1.75$ for the loose sand and $v_0 = 1.57$ for the dense sand.
The plastic hardening modulus is assigned as $h = 70$ for the loose specimen and $h = 120$ for the dense specimen.
The initial mean effective stress is $-390$ kPa for the loose case and $-425$ kPa for the dense case, with lateral earth pressure coefficients of $0.45$ and $0.38$, respectively.
\revised{These conditions result in constant radial stresses of -277.10 kPa (loose) and -275.28 kPa (dense), and initial axial stresses of -615.79 kPa (loose) and -724.43 kPa (dense), respectively.
The problem setup is shown in Fig.~\ref{fig:triaxial_setup}.}

\begin{figure}[h!]
    \centering
    \subfloat[Deviatoric stress--axial strain]{\includegraphics[width=0.45\textwidth]{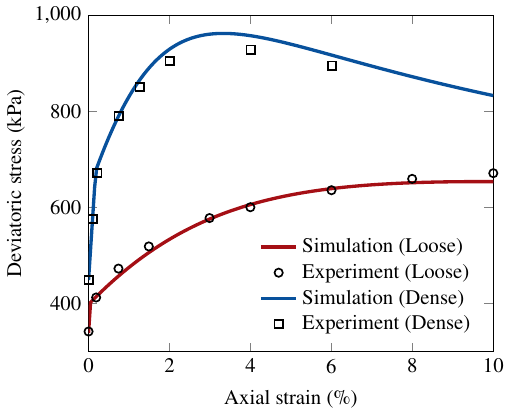}\label{fig:triaxial_test_simulation_results_dev}}\hspace{1em}
    \subfloat[Volumetric strain--axial strain]{\includegraphics[width=0.45\textwidth]{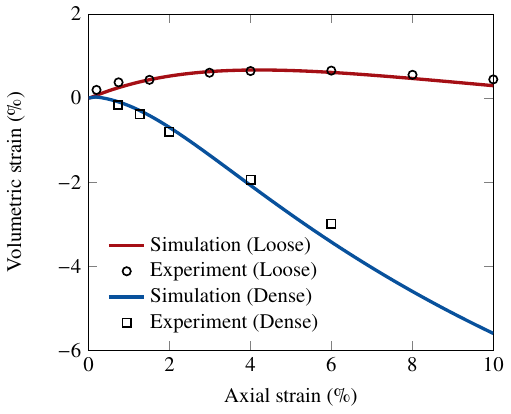}\label{fig:triaxial_test_simulation_results_vol}}
    \caption{Stress-point triaxial compression: simulation results obtained with the Nor-Sand model, compared against experimental data~\cite{cornforth1964some}.}
    \label{fig:triaxial_test_simulation_results}
\end{figure}

Figure~\ref{fig:triaxial_test_simulation_results} presents the simulation results of drained triaxial compression tests on Brasted sands under loose and dense conditions, modeled using the Nor-Sand model.
The deviatoric stress--axial strain responses are shown in Fig.~\ref{fig:triaxial_test_simulation_results_dev}, 
while the corresponding volumetric strain--axial strain curves are shown in Fig.~\ref{fig:triaxial_test_simulation_results_vol}. Experimental data from Conforth~\cite{cornforth1964some} are included for reference.
The simulation results closely reproduce the experimental responses, including the strain-softening behavior and volumetric contraction of loose sand, as well as the dilatant hardening behavior of dense sand. 
These results are consistent with the numerical results reported in Andrade and Ellison~\cite{andrade2008evaluation}, thereby verifying the model implementation.

\begin{figure}[h!]
    \centering
    \subfloat[Loose]{\includegraphics[width=0.45\textwidth]{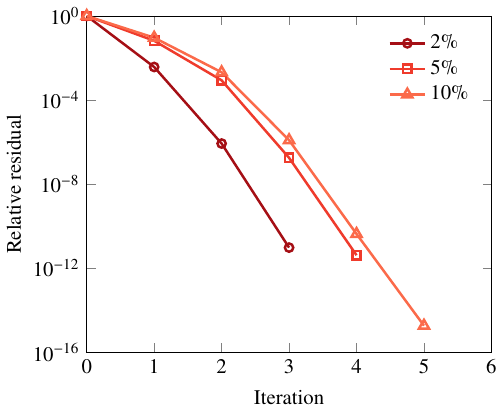}\label{fig:triaxial_test_convergence_loose}}{\hspace{1em}}
    \subfloat[Dense]{\includegraphics[width=0.45\textwidth]{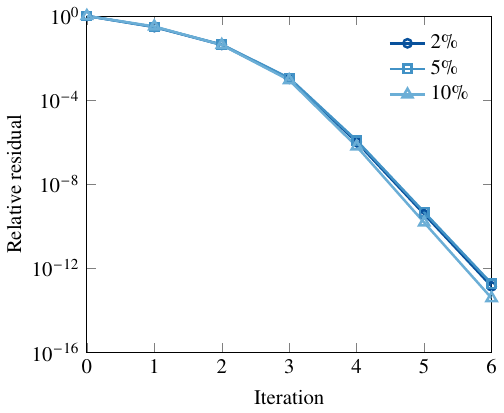}\label{fig:triaxial_test_convergence_dense}}
    \caption{Stress-point triaxial compression: convergence behavior of Newton iterations at selected axial strain levels for (a) the loose and (b) the dense Brasted sand, simulated using the Nor-Sand model.}
    \label{fig:triaxial_test_convergence}
\end{figure}

To evaluate the performance of AD in computing the consistent tangent operator, we examine the convergence behavior of Newton iterations used to control stress paths.
Figure~\ref{fig:triaxial_test_convergence} presents the residual norms at various axial strain levels for the loose and dense sands.
In all cases, the Newton solver converges within approximately five iterations for the loose sand and six for the dense sand.
The convergence rate is asymptotically quadratic, confirming that the consistent tangent operator is correctly computed via AD.
While Nor-Sand is one example, this AD-based algorithm generalizes to a wide range of complex, history-dependent constitutive models.
By enabling automatic and exact computation of consistent tangent operators, this methodology eliminates the need for manual derivation and implementation, thereby significantly streamlining the development, calibration, and validation of advanced constitutive models.

\subsection{Bar compaction under self-weight}
\begin{figure}[h!]
  \centering
  \includegraphics[width=0.24\textwidth]{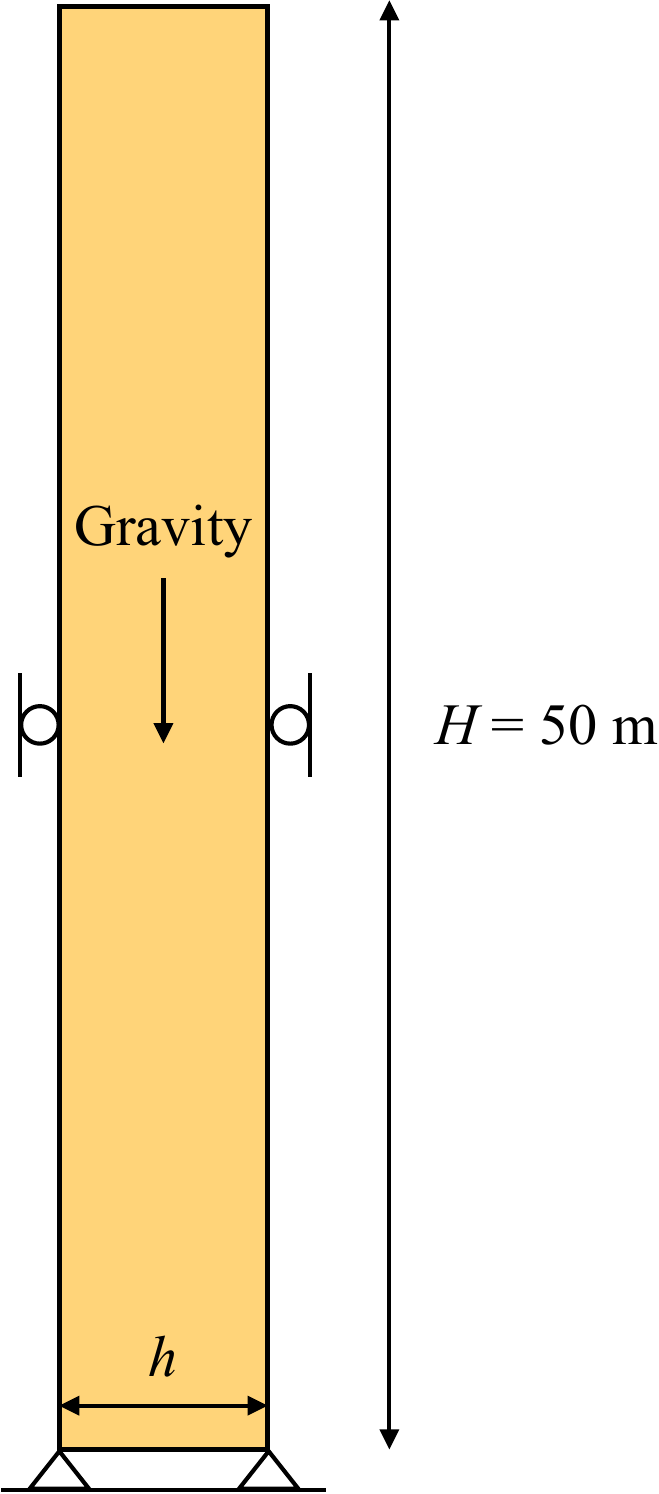}
  \caption{\revised{Bar compaction under self-weight: problem geometry and boundary conditions.}}
  \label{fig:bar_compaction_setup}
\end{figure}

The objective of the second example is to verify the formulation and implementation of MPM for solid mechanics.
To this end, we simulate the compaction of a vertical bar under self-weight—a standard benchmark problem in the MPM literature~\cite{charlton2017igimp,coombs2020ample,coombs2020lagrangian}.
\revised{Figure~\ref{fig:bar_compaction_setup} shows the problem setup.}
The bar has an initial height of $l_0 = 50$ m and an initial density of $\rho_0 = 80$ kg/m$^3$, and deforms under the action of gravity.
We consider two constitutive models: 
(1) elasticity based on Hencky strain (Hencky elasticity) and 
(2) elastoplasticity combining Hencky elasticity and J2 plasticity. 
Their material parameters are chosen such that the bar experiences large deformations during loading. 
For the elastic case, we assign a Young’s modulus of 10 kPa and a Poisson’s ratio of 0.
In the elastoplastic case where the yield surface is defined as $f(J_2, \kappa) = \sqrt{2 J_2} - \kappa \leq 0$ ($J_2$ is the second invariant of the deviatoric stress tensor), we assign a yield strength of $\kappa = 5$ kPa.
The elastic properties are taken to be identical to those in the purely elastic case.
Gravitational loading is applied incrementally over 40 load steps.

\begin{figure}[h!]
    \centering
    \subfloat[Elastic]{\includegraphics[width=0.45\textwidth]{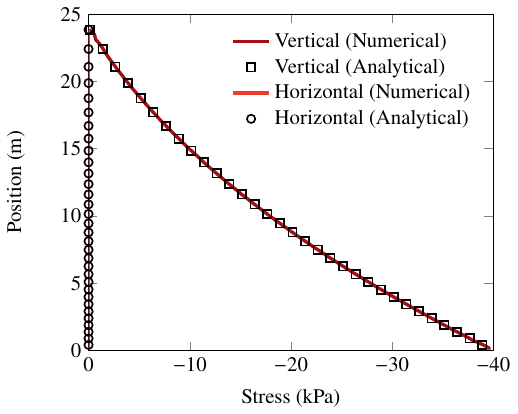}}\hspace{1em}
    \subfloat[Elastoplastic]{\includegraphics[width=0.45\textwidth]{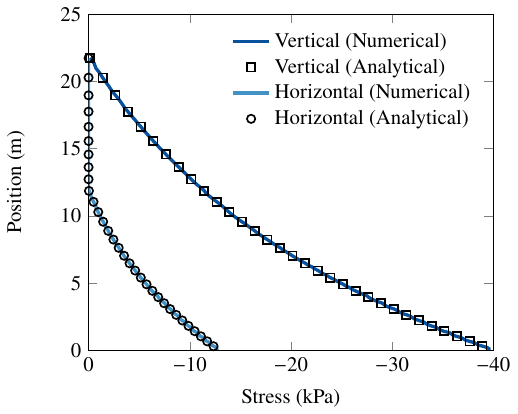}}
    \caption{Bar compaction under self-weight: position--stress distributions for the (a) elastic and (b) elastoplastic case.}
    \label{fig:bar_compaction_position_stress}
\end{figure}

Figure~\ref{fig:bar_compaction_position_stress} shows the final vertical and horizontal stress distributions along the bar, computed using a background grid size of $h = 50/64$ m and 256 particles.
The analytical solutions for both the elastic and elastoplastic cases are taken from Charlton~\etal~\cite{charlton2017igimp} and plotted for comparison.
The final height of the bar is observed to be less than 25 m, indicating substantial compaction from the initial height of 50 m.
The numerical results closely agree with the analytical solutions in both cases, thereby confirming the accuracy of the MPM formulation and its implementation.

\begin{figure}[h!]
    \centering
    \subfloat[Elastic]{\includegraphics[width=0.45\textwidth]{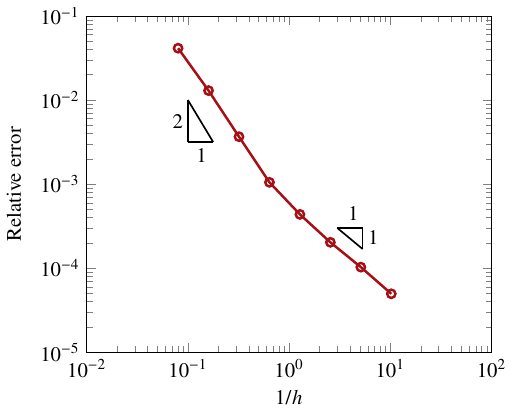}}\hspace{1em}
    \subfloat[Elastoplastic]{\includegraphics[width=0.45\textwidth]{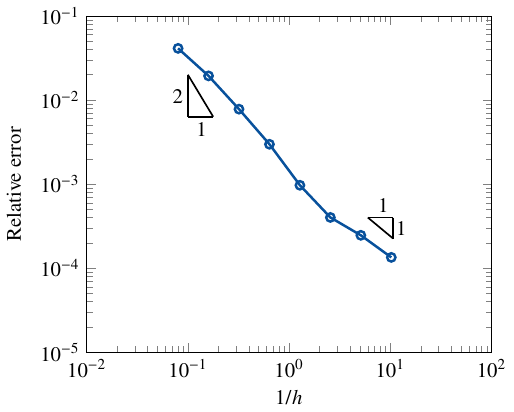}}
    \caption{Bar compaction under self-weight: convergence of the relative error with spatial refinement for the (a) elastic and (b) elastoplastic cases.}
    \label{fig:bar_compaction_mesh_convergence}
\end{figure}

To further evaluate accuracy, we conduct a convergence study using eight levels of spatial discretization.
The number of background cells in the vertical direction ranges from $2^2$ to $2^9$, with the corresponding number of particles varying from 16 to 2048.
Figure~\ref{fig:bar_compaction_mesh_convergence} shows the convergence of the relative error with mesh refinement, where the relative error is defined as
\begin{equation}
    \text{error} = \sum_p \dfrac{\Vert \sigma_{p,yy} - \sigma_{a,yy} \Vert V_p^0}{g \rho_0 l_0 V_p^0},
\end{equation}
with $\sigma_{p,yy}$ denoting the numerical vertical stress and $\sigma_{a,yy} = \rho_0 g (l_0 - Y)$ the analytical solution at height $Y$.
In both the elastic and elastoplastic cases, the numerical results exhibit a convergence rate between 1 and 2, consistent with prior observations reported in the literature~\cite{charlton2017igimp,coombs2020ample}.

\begin{figure}[h!]
    \centering
    \subfloat[Elastic]{\includegraphics[width=0.45\textwidth]{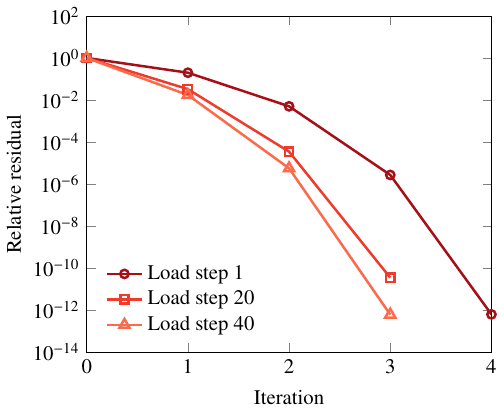}}\hspace{1em}
    \subfloat[Elastoplastic]{\includegraphics[width=0.45\textwidth]{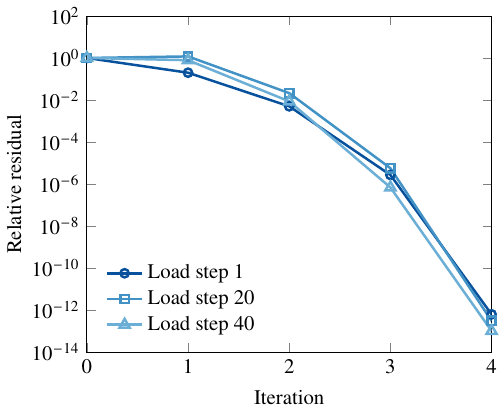}}
    \caption{Bar compaction under self-weight: relative residual norms during global Newton iterations at selected load steps for the (a) elastic and (b) elastoplastic cases.}
    \label{fig:bar_compaction_newton_convergence}
\end{figure}

In addition to spatial convergence, we examine the convergence behavior of the global Newton iterations.
Figure~\ref{fig:bar_compaction_newton_convergence} shows the residual norm versus iteration count at three representative load steps.
In both the elastic and elastoplastic cases, the iterations exhibit asymptotically quadratic convergence.
As expected, the elastoplastic case requires slightly more iterations than the elastic case due to increased nonlinearity.
Nonetheless, all iterations converge to a residual tolerance of $10^{-11}$ within four steps.
These results verify the MPM implementation, demonstrating accurate spatial convergence and robust Newton solver performance.

\subsection{Cantilever beam}
\begin{figure}[h!]
    \centering
    \includegraphics[width=0.75\textwidth]{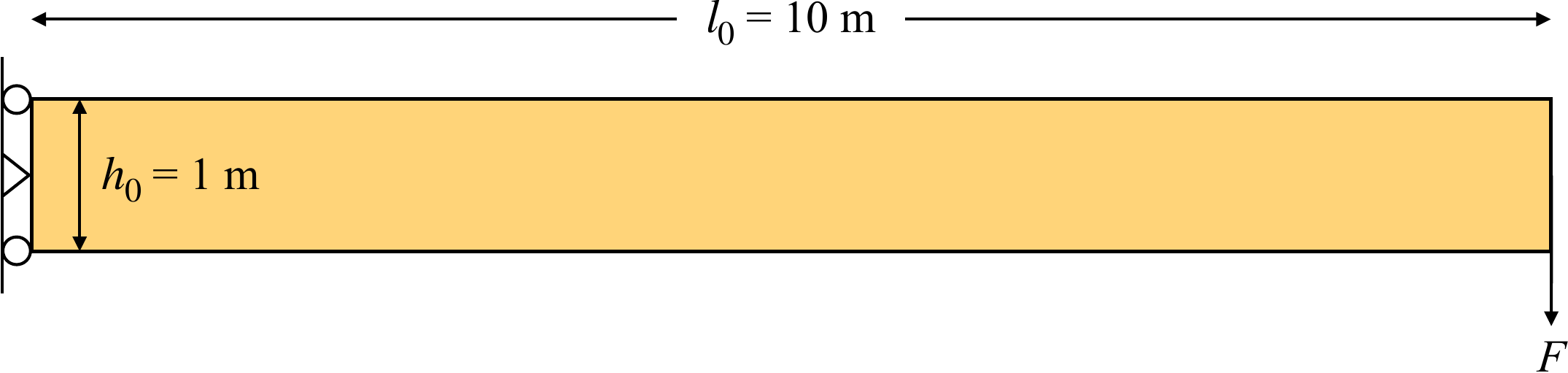}
    \caption{Cantilever beam: problem geometry and boundary conditions.}
    \label{fig:cantilever_beam_setup}
\end{figure}

The third example evaluates the efficiency of the proposed sparse automatic differentiation algorithm for Jacobian construction in MPM simulations, while also providing further verification of the formulation.
To this end, we simulate the bending of a cantilever beam, which is a standard benchmark problem for verifying MPM formulations~\cite{charlton2017igimp,coombs2020ample,coombs2020lagrangian}.
The problem setup is illustrated in Fig.~\ref{fig:cantilever_beam_setup}, where a point load of $F = 100$ kN is applied at the free end of the beam.
The beam material is modeled using Hencky elasticity with a Young’s modulus of 12 MPa and a Poisson’s ratio of 0.2.
The load is applied incrementally over 50 steps.

\begin{figure}[h!]
\centering
\subfloat[Load step 4]{\includegraphics[width=0.45\textwidth]{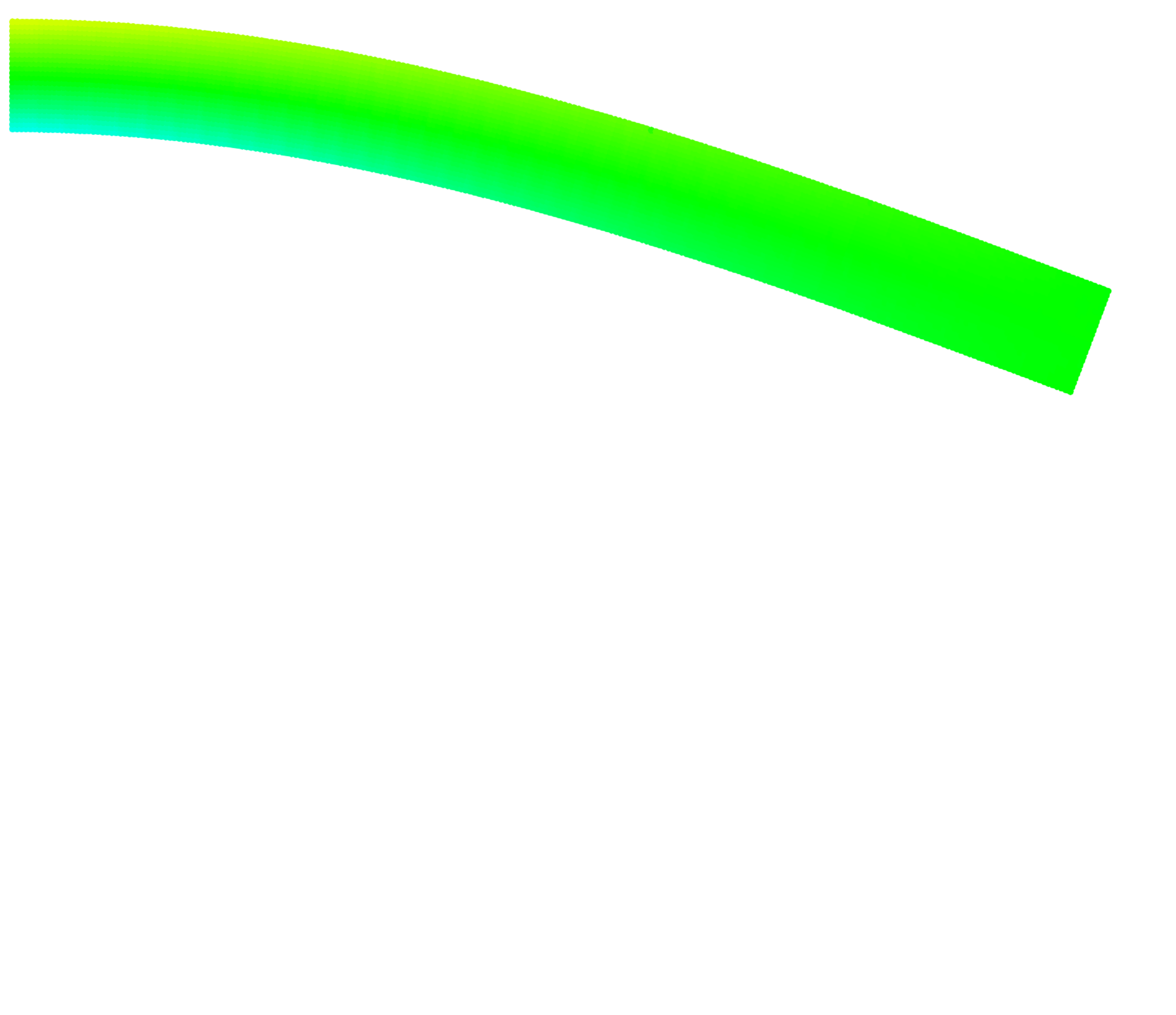}}\hspace{1em}
\subfloat[Load step 8]{\includegraphics[width=0.45\textwidth]{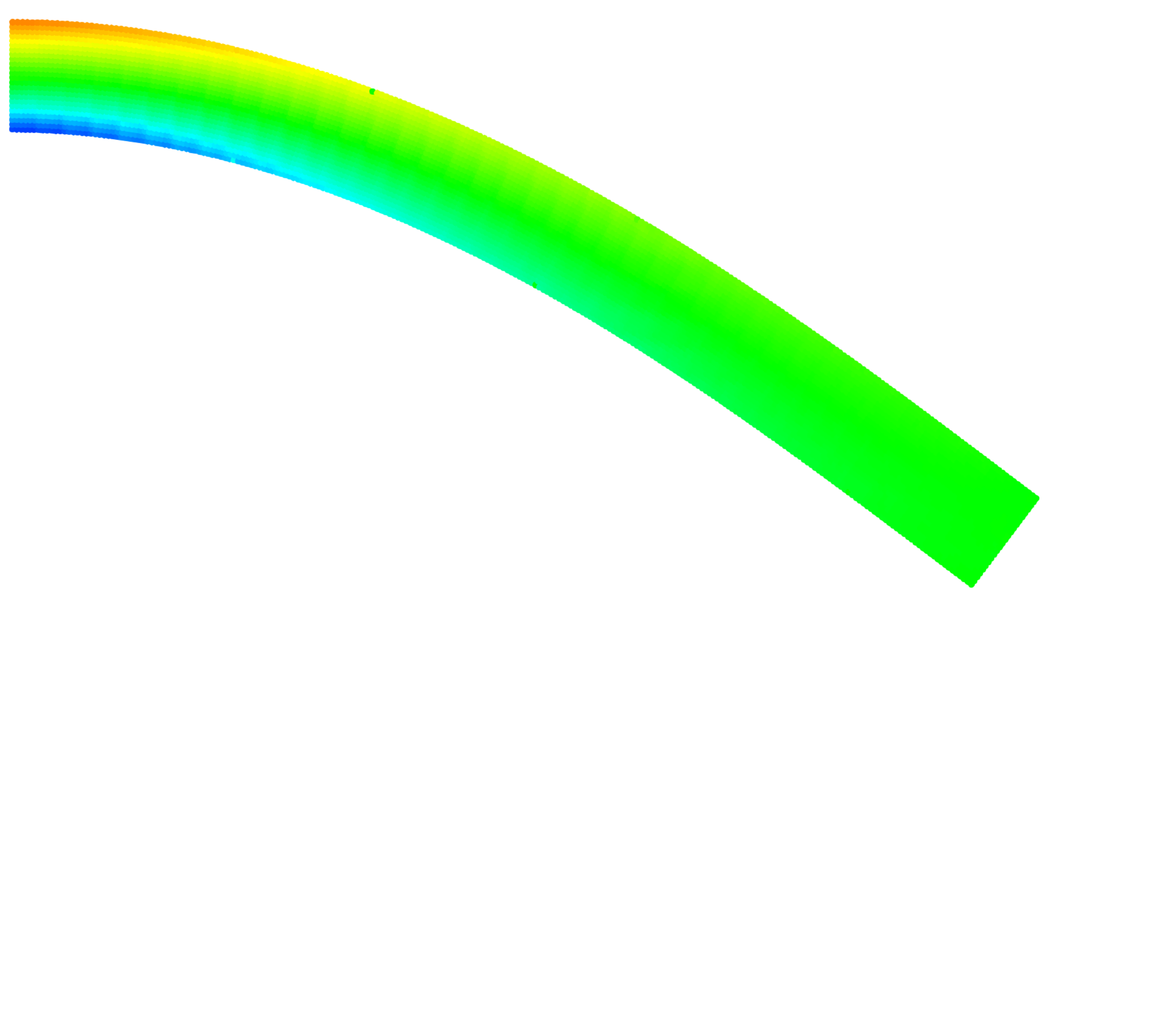}}\
\subfloat[Load step 16]{\includegraphics[width=0.45\textwidth]{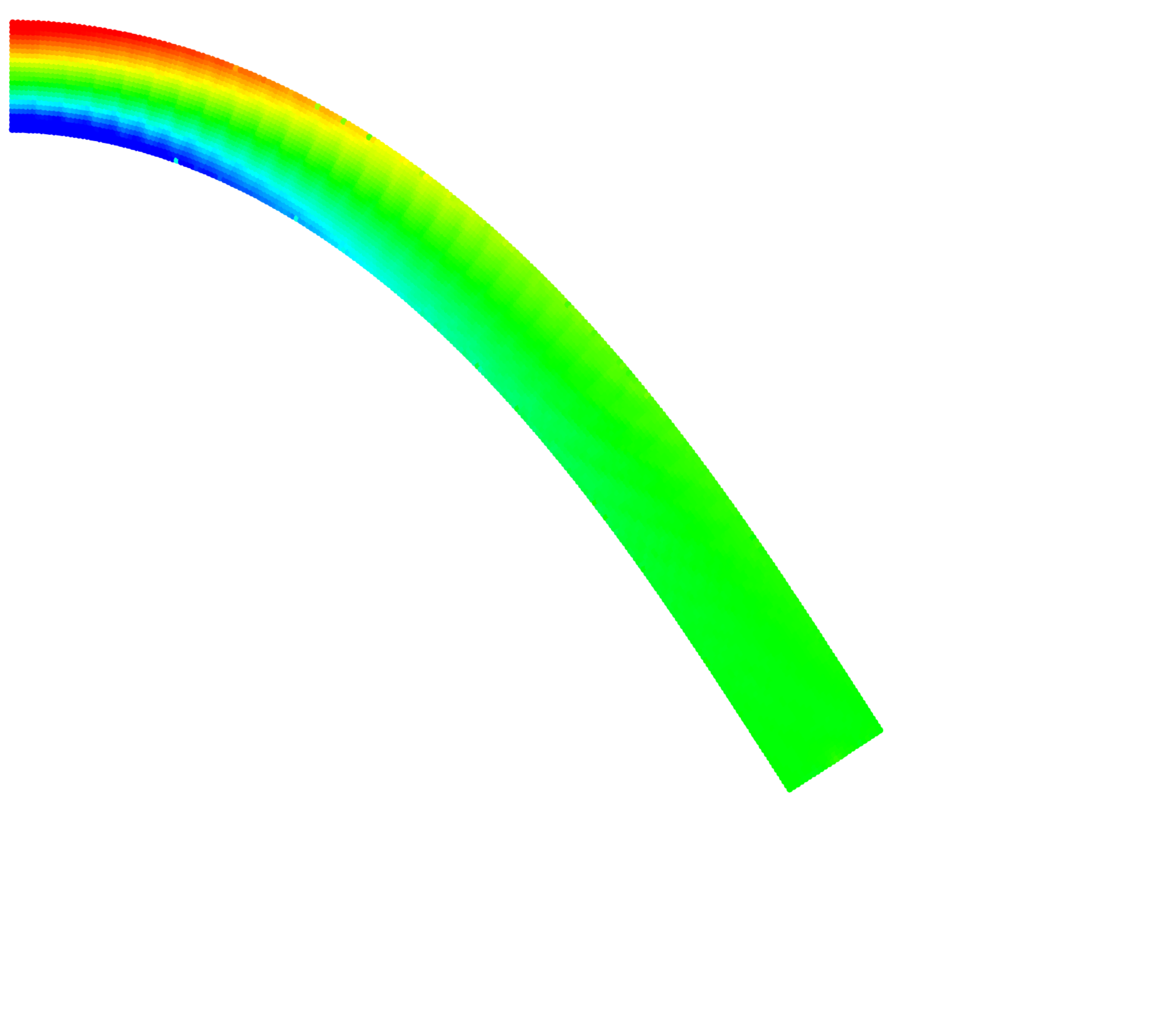}}\hspace{1em}
\subfloat[Load step 50]{\includegraphics[width=0.45\textwidth]{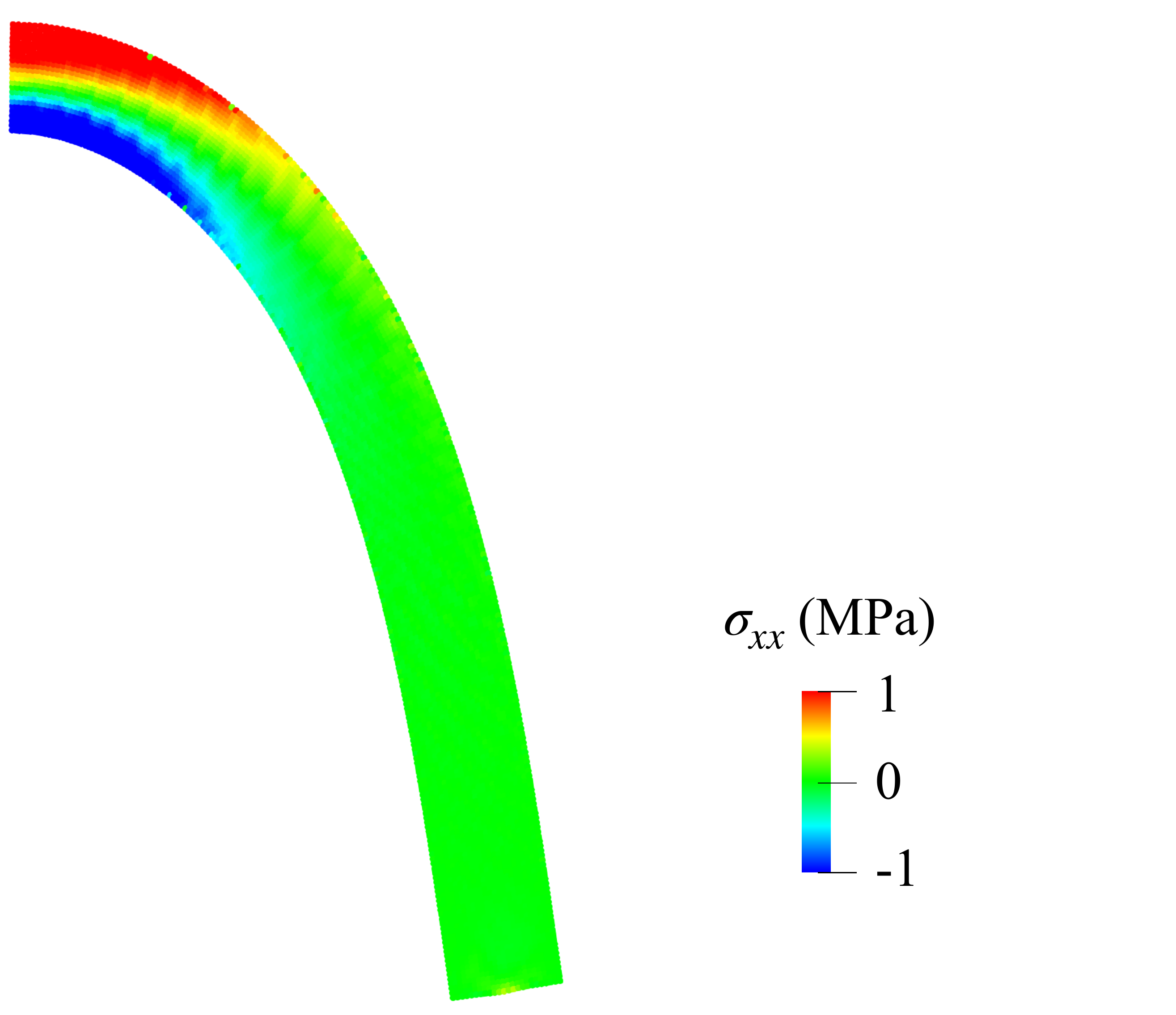}}
\caption{Cantilever beam: deformed configurations colored by horizontal stress at (a) load step 4, (b) load step 8, (c) load step 16, and (d) load step 50. The background grid size is $h = 0.25$ m.}
\label{fig:cantilever_beam_deformed_configuration}
\end{figure}

Figure~\ref{fig:cantilever_beam_deformed_configuration} shows the deformed configurations of the cantilever beam at selected load steps, computed using a background grid size of $h = 0.25$ m with 5,760 particles (corresponding to $6 \times 6$ particles per cell). As the applied load increases, the beam undergoes progressively larger deformations, which are accurately captured by the simulation.

\begin{figure}[h!]
\centering
\subfloat[Vertical]{\includegraphics[width=0.45\textwidth]{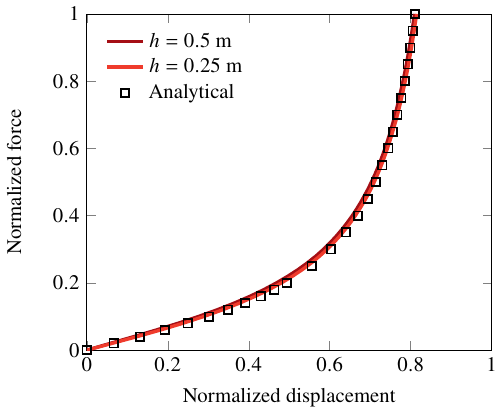}}\hspace{1em}
\subfloat[Horizontal]{\includegraphics[width=0.45\textwidth]{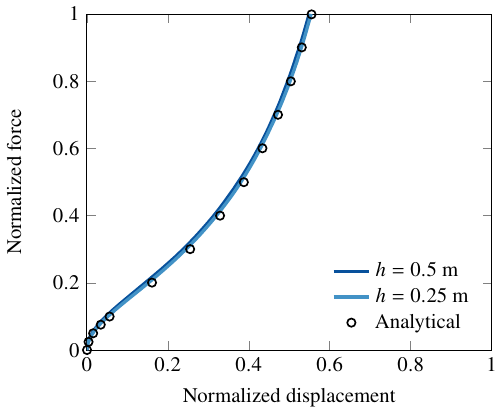}}
\caption{Cantilever beam: normalized force–displacement responses in the (a) vertical and (b) horizontal directions, compared against analytical solutions~\cite{molstad1977finite}.}
\label{fig:cantilever_beam_verification}
\end{figure}

The simulation results are further verified against analytical solutions derived by Molstad~\cite{molstad1977finite}. As shown in Fig.~\ref{fig:cantilever_beam_verification}, the normalized vertical and horizontal force--displacement responses exhibit excellent agreement with the analytical solutions across two levels of spatial discretization, confirming the accuracy of the formulation.

\begin{figure}[h!]
  \centering
  \includegraphics[width=0.5\textwidth]{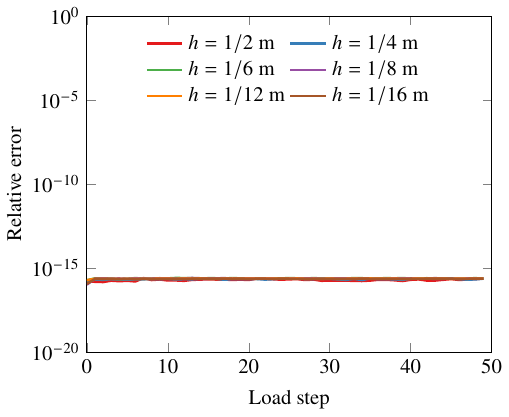}
  \caption{\revised{Cantilever beam: relative error of the Jacobian matrices obtained by the sparse and dense differentiation algorithms.}}
  \label{fig:cantilever_beam_jacobian}
\end{figure}

\begin{figure}[h!]
    \centering
    \includegraphics[width=0.5\textwidth]{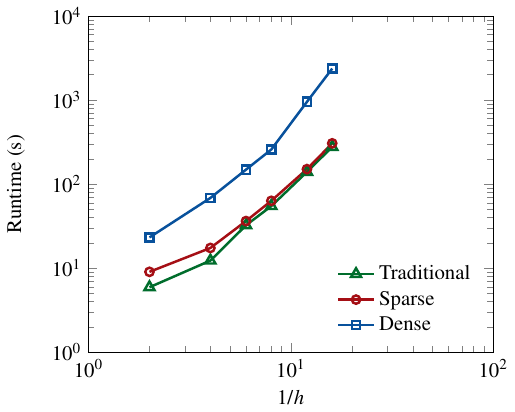}
    \caption{\revised{Cantilever beam: total runtime comparison between the sparse and dense differentiation algorithms for Jacobian construction, along with the traditional differentiation algorithm without AD.}}
    \label{fig:cantilever_beam_total_runtime}
\end{figure}

\revised{Next, we evaluate the accuracy and computational efficiency of the proposed sparse differentiation algorithm for Jacobian construction.}
Six levels of discretization are tested, with background grid sizes and corresponding particle counts as follows:
(1) $h = 1/2$ m (1,440 particles), (2) $h = 1/4$ m (5,760 particles), (3) $h = 1/6$ m (12,960 particles), (4) $h = 1/8$ m (23,040 particles), (5) $h = 1/12$ m (51,840 particles), and (6) $h = 1/16$ m (92,160 particles).
\revised{To verify that sparse differentiation yields numerically identical results to dense differentiation, we directly compare the Jacobian matrices produced by the two methods.
Figure~\ref{fig:cantilever_beam_jacobian} shows the relative error in the Frobenius norm---a commonly used matrix norm~\cite{golub2013matrix}---across all discretization levels.
The error remains below $10^{-15}$, confirming that the sparse differentiation does not introduce additional truncation errors. 
To assess computational efficiency, we compare the total simulation runtime of three methods: the traditional differentiation without AD, the proposed sparse differentiation, and the standard dense differentiation.
As shown in Fig.~\ref{fig:cantilever_beam_total_runtime}, the sparse method achieves performance nearly identical to the traditional non-AD differentiation, while the dense method is significantly slower at all resolutions.
At the finest grid, the sparse method achieves nearly an eightfold speedup in Jacobian construction compared to the dense implementation.}

\begin{figure}[h!]
    \centering
    \includegraphics[width=0.5\textwidth]{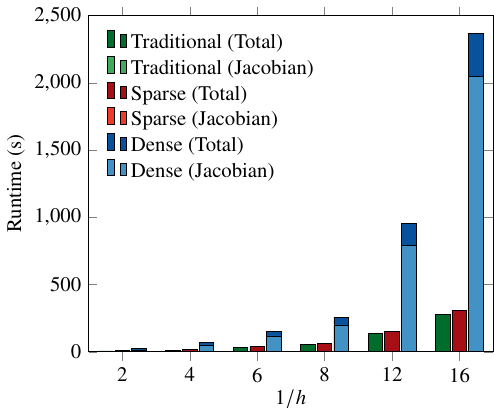}
    \caption{\revised{Cantilever beam: comparison of computational time for the sparse, dense, and traditional algorithms used in Jacobian construction.}}
    \label{fig:cantilever_beam_comparison_of_time}
\end{figure}

To further evaluate the computational savings, we isolate the time spent specifically on Jacobian \revised{assembly}.
The timing results for \revised{the three} algorithms are summarized in Table~\ref{tab:cantilever_beam_breakdown} and visualized in Fig.~\ref{fig:cantilever_beam_comparison_of_time}.
\revised{As shown, both the traditional and sparse differentiation algorithms exhibit nearly constant assembly time across all discretization levels.
The sparse differentiation is slightly slower because of the overhead from AD.}
\revised{Nevertheless, this cost for differentiation becomes negligible relative to the total runtime, as seen in Fig.~\ref{fig:cantilever_beam_comparison_of_time}.}
In contrast, the \revised{assembly} step in the dense differentiation dominates the total simulation time, accounting for roughly 85\% of the total runtime at the finest resolution.
\begin{table}[h!]
    \centering
    \begin{tabular}{ccccccc}
    \hline
    \revised{Grid size} (m) & \multicolumn{2}{c}{\revised{Traditional} (s)} & \multicolumn{2}{c}{\revised{Sparse} (s)} & \multicolumn{2}{c}{\revised{\revised{Dense} (s)}} \\
    \cline{2-3} \cline{4-5} \cline{6-7}
    & \revised{Total} & \revised{Jacobian} & \revised{Total} & \revised{Jacobian} & \revised{Total} & \revised{Jacobian} \\
    \hline
    $1/2$ & \revised{5.92} & \revised{0.73 (12.33\%)}  & 9.03   & 2.73 (30.25\%)  & 23.18   & 12.94 (55.85\%)   \\
    $1/4$ & \revised{12.29} & \revised{0.69 (5.61\%)}   & 17.37  & 2.73 (15.72\%)  & 68.38   & 50.68 (74.11\%)  \\
    $1/6$ & \revised{32.34} & \revised{0.74 (2.29\%)}  & 36.30  & 2.78 (7.67\%)   & 148.53  & 117.70 (79.24\%)  \\
    $1/8$ & \revised{54.69} & \revised{0.83 (1.52\%)} & 63.17  & 2.89 (4.58\%)   & 257.65  & 199.27 (77.34\%)  \\
    $1/12$ & \revised{139.45} & \revised{0.86 (0.62\%)} & 150.42 & 2.92 (1.94\%)   & 951.06  & 792.37 (83.31\%)  \\
    $1/16$ & \revised{277.12} & \revised{0.75 (0.27\%)} & 305.13 & 3.02 (0.99\%)   & 2366.15 & 2046.43 (86.49\%) \\
    \hline
    \end{tabular}
    \caption{\revised{Cantilever beam: breakdown of assembly time for the sparse, dense, and traditional algorithms for Jacobian construction.}}
    \label{tab:cantilever_beam_breakdown}
\end{table}

\begin{figure}[h!]
  \centering
  \subfloat[Roofline]{\includegraphics[width=0.45\textwidth]{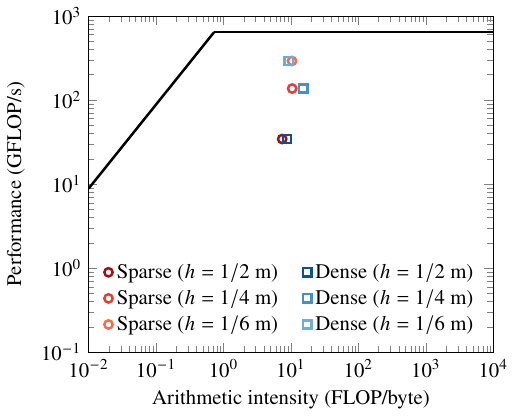}\label{fig:cantilever_beam_roofline_performance}}\hspace{1em}
  \subfloat[Work]{\includegraphics[width=0.45\textwidth]{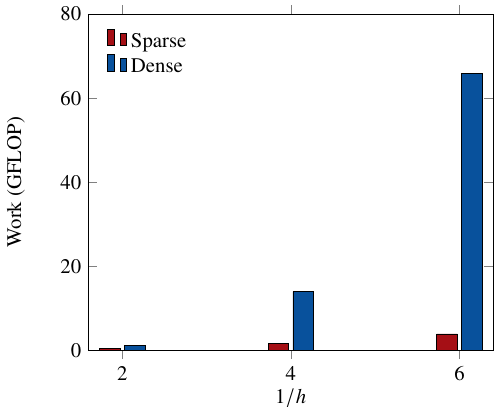}\label{fig:cantilever_beam_roofline_work}}
  \caption{\revised{Cantilever beam: (a) Roofline analysis showing performance versus arithmetic intensity, and (b) total work (GFLOPs) versus spatial refinement for sparse and dense differentiation algorithms.}}
  \label{fig:cantilever_beam_roofline}
\end{figure}

\revised{We also conduct a Roofline analysis on an NVIDIA GeForce RTX 5070 Ti GPU using NVIDIA Nsight Compute to assess the performance characteristics of the proposed algorithm. The objective is to determine whether the sparse and dense implementations are memory-bound or compute-bound. As shown in Fig.~\ref{fig:cantilever_beam_roofline_performance}, both algorithms occupy similar positions near the compute-bound region, indicating that performance is limited by floating-point throughput rather than memory bandwidth. The similarity between the two further suggests that the speedup observed in the sparse formulation arises primarily from a reduction in the number of AD backward passes, rather than improved per-kernel efficiency. In other words, sparse differentiation accelerates computation by reducing the total workload while preserving similar arithmetic intensity per kernel. This interpretation is supported by Fig.~\ref{fig:cantilever_beam_roofline_work}, which shows that the sparse differentiation executes significantly fewer GFLOPs than the dense differentiation. Notably, the gap between the two widens with increasing resolution, indicating that the work savings achieved by sparse differentiation become more substantial as the grid is refined. This trend underscores the scalability of the sparse formulation for large-scale problems.}

These results demonstrate that the proposed sparse differentiation method significantly reduces computational overhead for Jacobian construction, especially in simulations with large numbers of degrees of freedom.
Such efficiency gains are critical for practical applications of MPM with automatic differentiation, where fine spatial discretization is often required.

\subsection{1D consolidation}
\begin{figure}[h!]
    \centering
    \includegraphics[width=0.24\textwidth]{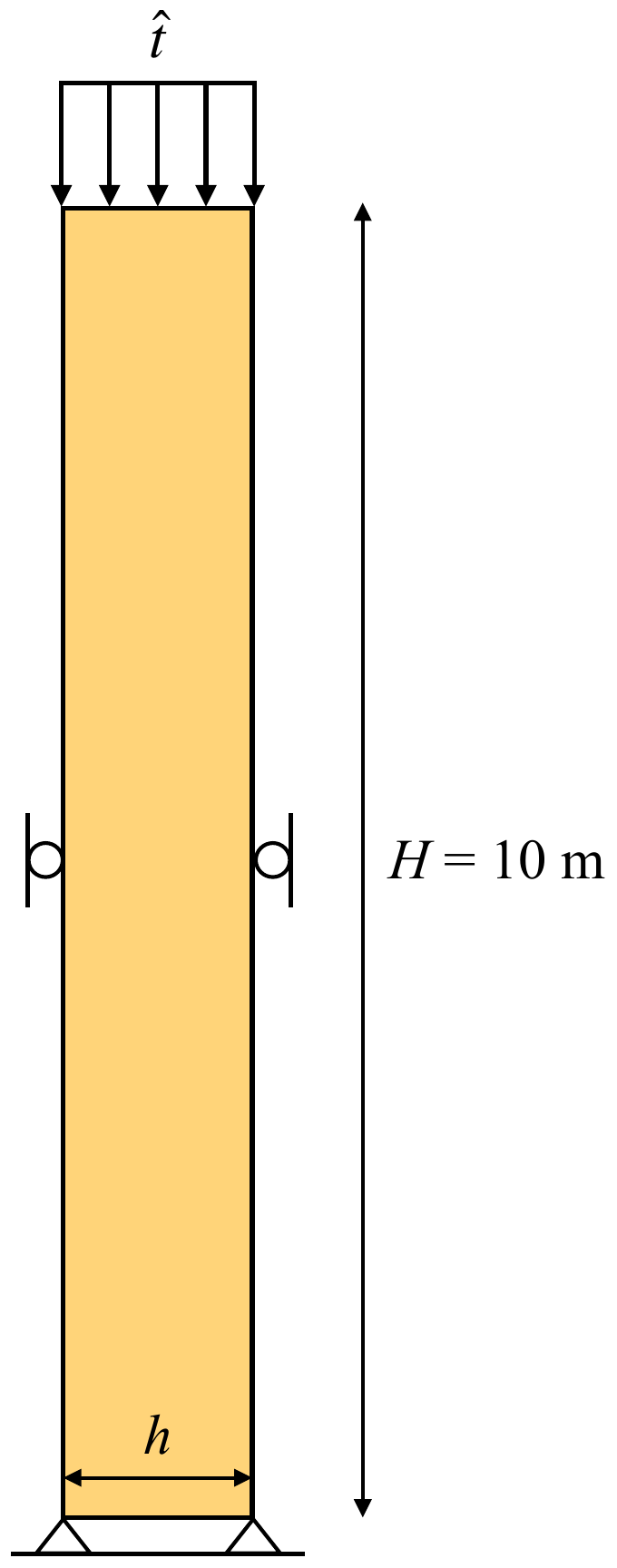}
    \caption{1D consolidation: problem geometry and boundary conditions.}
    \label{fig:consolidation_setup}
\end{figure}

The fourth example verifies the implementation of the coupled $\tensor{u}$–$p$ formulation, which is widely employed in geomechanical simulations of fluid-saturated porous media.
To this end, we simulate 1D consolidation behavior under both small and large deformation conditions.
The problem geometry and boundary conditions are shown in Fig.~\ref{fig:consolidation_setup}. 
A 10-meter-tall porous column is subjected to a uniformly distributed surface load of $\hat{t} = 1$ kPa applied at the top, which also serves as the drainage boundary.

For the solid skeleton, we adopt a Neo-Hookean hyperelastic model.
The Lam\'{e} parameters are set to $\lambda = 600$ kPa and $\mu = 600$ kPa.
Fluid flow is governed by Darcy’s law. 
\revised{To demonstrate the capability of the proposed framework in simulating long-term processes, a small constant intrinsic permeability of $k = 10^{-15}$~m$^2$ is assigned. The fluid density is set to $\rho = 1$~t/m$^3$.}
These parameters give a coefficient of consolidation of \revised{$c_v = 1.8 \times 10^{-6}$} m$^2$/s.
The computational domain is discretized using 100 background grid cells in the vertical direction and one cell in the horizontal direction, with $2 \times 2$ particles per cell.
\revised{To simulate this long-term behavior within a reasonable time, the time step is set to $\Delta t = 10^5$ s.}

\begin{figure}[h!]
    \centering
    \includegraphics[width=0.5\textwidth]{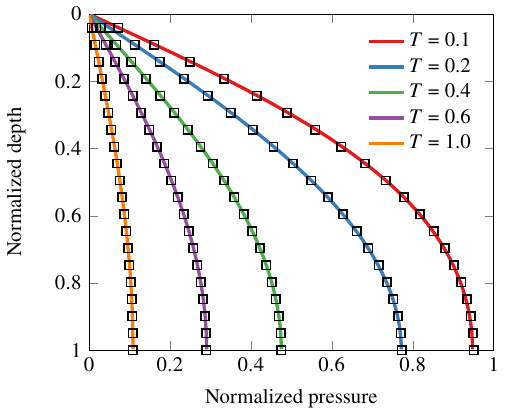}
    \caption{\revised{1D consolidation: pore pressure distributions at selected nondimensional times compared with analytical solutions based on Terzaghi’s analytical solution.}}
    \label{fig:consolidation_terzaghi}
\end{figure}

Figure~\ref{fig:consolidation_terzaghi} shows the simulated pore pressure distributions at selected time instances, compared against Terzaghi’s analytical solution for one-dimensional consolidation under infinitesimal strain assumptions.
\revised{The simulation is carried out up to a nondimensional time factor of $T := c_v t / H^2 = 1$, which corresponds to a physical duration of approximately 640 days.}
The numerical results exhibit excellent agreement with the analytical profiles at all time steps, confirming the accuracy of the coupled $\tensor{u}$--$p$ formulation in the infinitesimal deformation regime.
\revised{These results demonstrate that GeoWarp can robustly simulate long-term consolidation processes.}

To further verify the formulation under large deformation conditions, we modify the problem setup following Uzuoka and Borja~\cite{uzuoka2012dynamics}.
Specifically, the surface load is increased to $\hat{t}=10$ kPa, and the Lam\'{e} parameters are reduced to $\lambda = 10$ kPa and $\mu = 15$ kPa to model a softer solid skeleton.
The initial intrinsic permeability is increased to $k_0 = 10^{-9}$ m$^2$ and evolves with deformation according to the constitutive relation described in Uzuoka and Borja~\cite{uzuoka2012dynamics}.
The initial time step is set to $\Delta t = 0.1$ s and is increased by a factor of 1.2 at each step, up to a maximum of 80 s. 
Under large deformation, applying pore pressure boundary conditions at fixed grid nodes can lead to abrupt changes in the drainage boundary.
To address this, we implement a moving mesh strategy following the algorithm in Al-Kafaji~\cite{alkafaji2013formulation}, wherein the background grid deforms with the consolidating domain to maintain consistent boundary enforcement.

\begin{figure}[h!]
    \centering
    \includegraphics[width=0.5\textwidth]{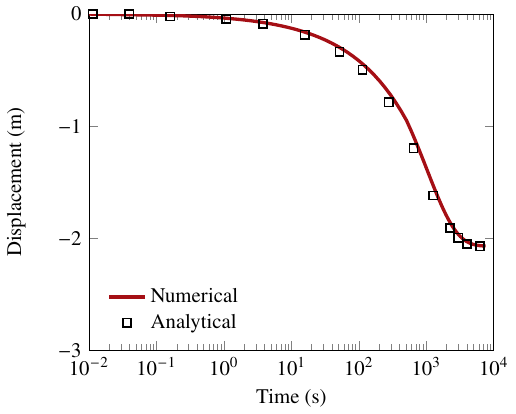}
    \caption{1D consolidation: evolution of top surface displacement over time, compared against the analytical solution at finite strains~\cite{uzuoka2012dynamics}.}
    \label{fig:consolidation_finite}
\end{figure}

Figure~\ref{fig:consolidation_finite} shows the evolution of the top surface displacement over time, compared against the analytical solution at finite strains presented in Uzuoka and Borja~\cite{uzuoka2012dynamics}.
The numerical solution shows excellent agreement with the analytical solution in both the transient and steady-state responses,  confirming that the coupled $\tensor{u}$--$p$ formulation is correctly implemented even under large deformation conditions.

Importantly, the coupled $\tensor{u}$--$p$ formulation is fully compatible with automatic differentiation.
This feature not only enables automatic computation of the Jacobian matrix but also facilitates gradient-based optimization.
The latter capability is demonstrated in the following example.

\subsection{Ground stiffness identification via inverse analysis}
\begin{figure}[h!]
    \centering
    \includegraphics[width=0.5\textwidth]{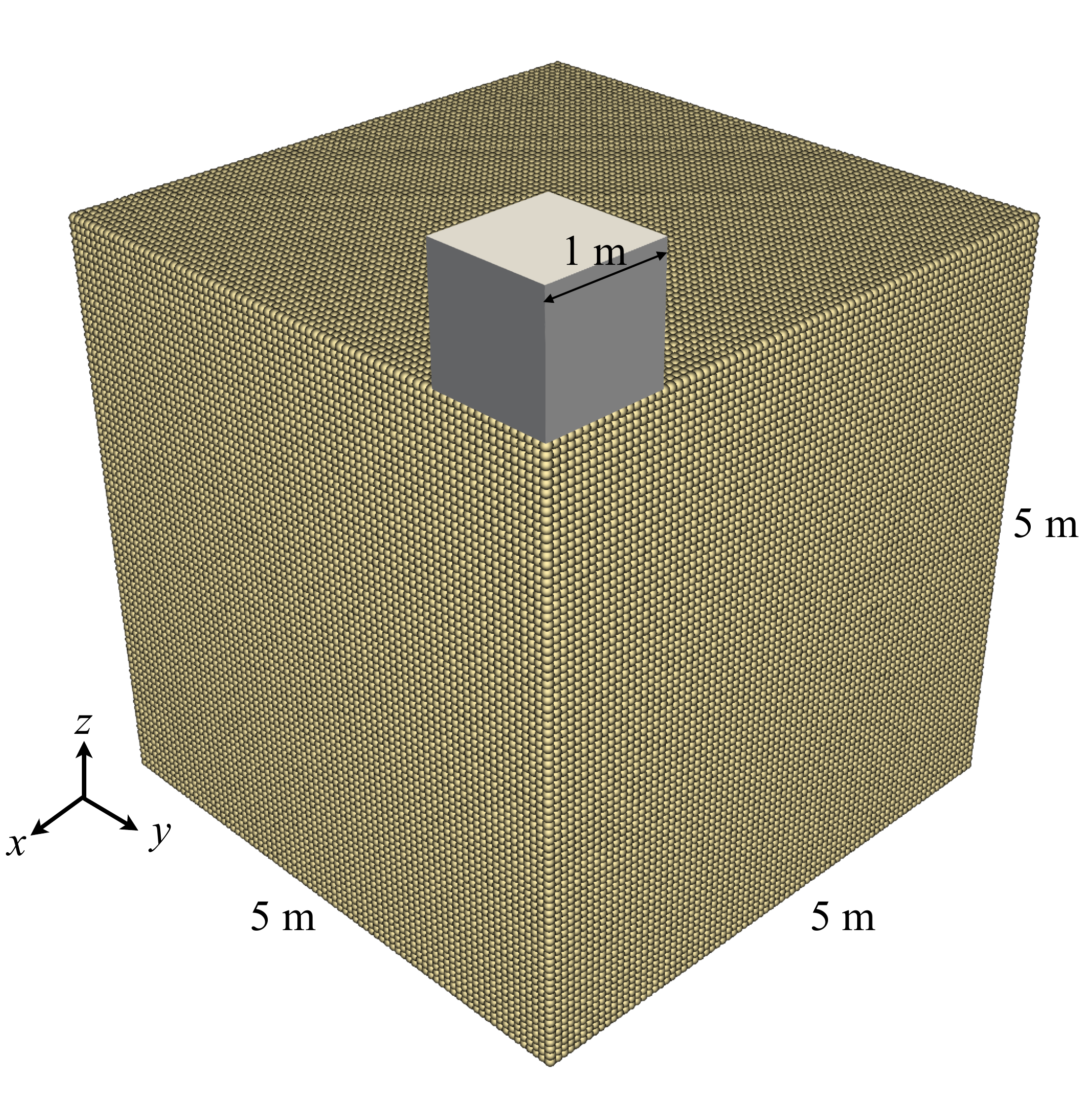}
    \caption{Ground stiffness identification via inverse analysis: problem geometry and boundary conditions.}
    \label{fig:indentation_setup}
\end{figure}

The fifth and final example demonstrates GeoWarp's capability to perform inverse analysis using automatic differentiation. 
Specifically, we consider the identification of ground stiffness from the indentation response of a rigid footing into saturated porous ground.
The problem geometry and boundary conditions are illustrated in Fig.~\ref{fig:indentation_setup}. A 5 m $\times$ 5 m domain of saturated porous ground is indented by a rigid footing of size 1 m $\times$ 1 m. 
Horizontal displacements are constrained along the lateral boundaries in both the $x$- and $y$-directions, while the bottom boundary is fixed in the $z$-direction. 
The top surface is modeled as a drainage boundary.

As in the previous example, the ground is modeled as a Neo-Hookean hyperelastic material.
\revised{While this choice is common for large-deformation problems, it is adopted here primarily to enable gradient-based inverse analysis, as a hyperelastic model ensures smooth and differentiable constitutive behavior throughout the simulation. Although elastoplastic soil models can be used in forward analysis, they introduce non-differentiability due to elastic–plastic transitions and nonsmooth yield surfaces such as Mohr–Coulomb, which complicates gradient-based methods. While the framework supports automatic differentiation, it cannot eliminate non-differentiability inherent to the constitutive model. For this reason, we focus on an elastic material to clearly demonstrate the inverse analysis capabilities of GeoWarp under differentiable conditions.}
As for the material parameters, the Young’s modulus is set to $E = 10$ MPa, and the Poisson’s ratio to $\nu = 0.3$. The intrinsic permeability is assumed constant at $k = 10^{-14}$ m$^2$. The domain is discretized using a uniform background grid of size 0.25 m $\times$ 0.25 m $\times$ 0.25 m, with $4 \times 4 \times 4$ particles per cell, resulting in a total of 512,000 particles. 
The time increment is fixed at $\Delta t = 0.1$ s. 
Contact between the rigid footing and the porous ground is modeled using a penalty-based method~\cite{chandra2021nonconforming}, with a penalty factor of 1,000. 
To prevent abrupt changes in pore pressure boundaries under large deformation, a moving mesh strategy is employed following Al-Kafaji~\cite{alkafaji2013formulation}.

\begin{figure}[h!]
    \centering
    \subfloat[Load step 10]{\includegraphics[height=0.43\textwidth]{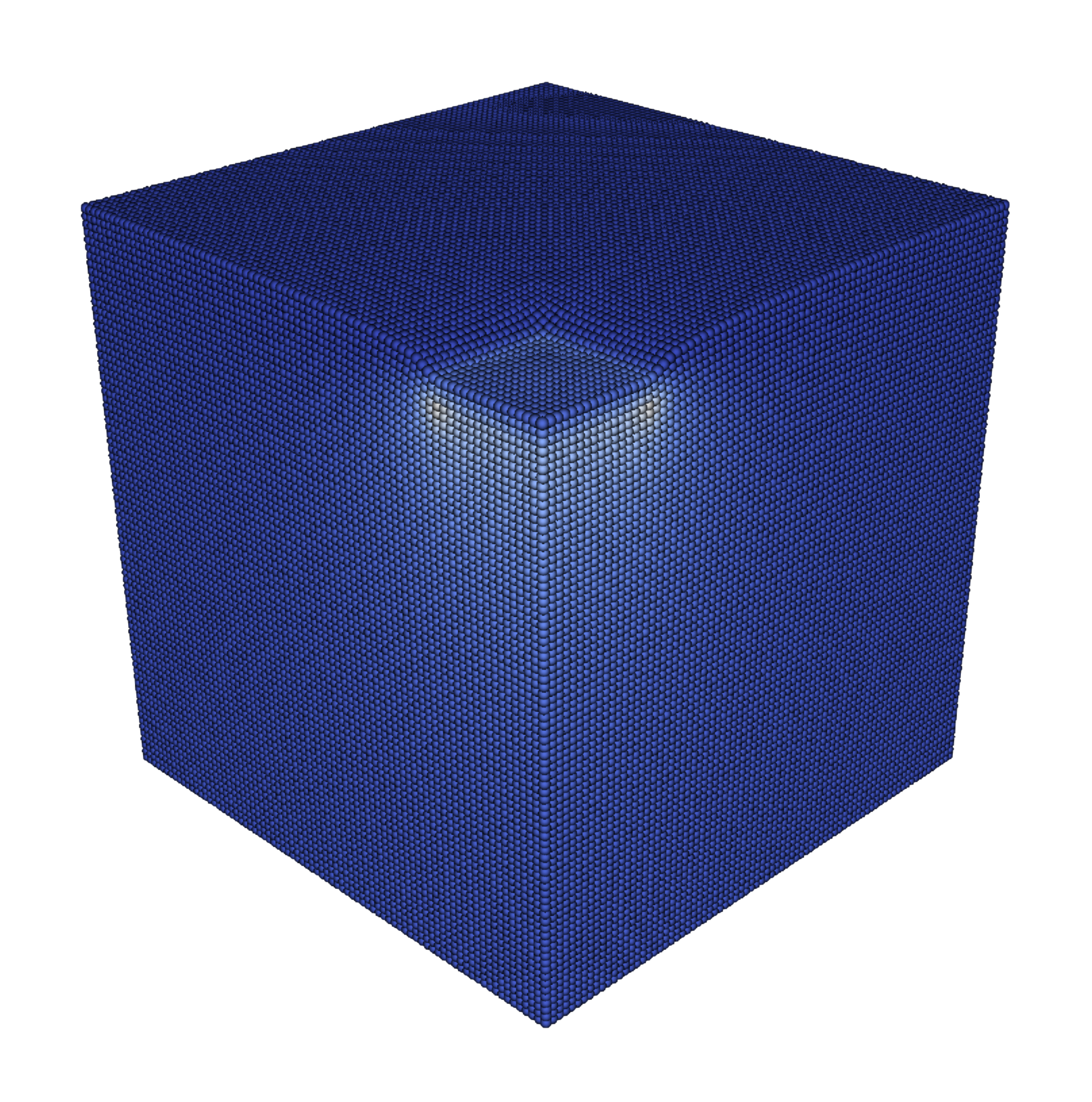}}\hspace{1em}
    \subfloat[Load step 15]{\includegraphics[height=0.43\textwidth]{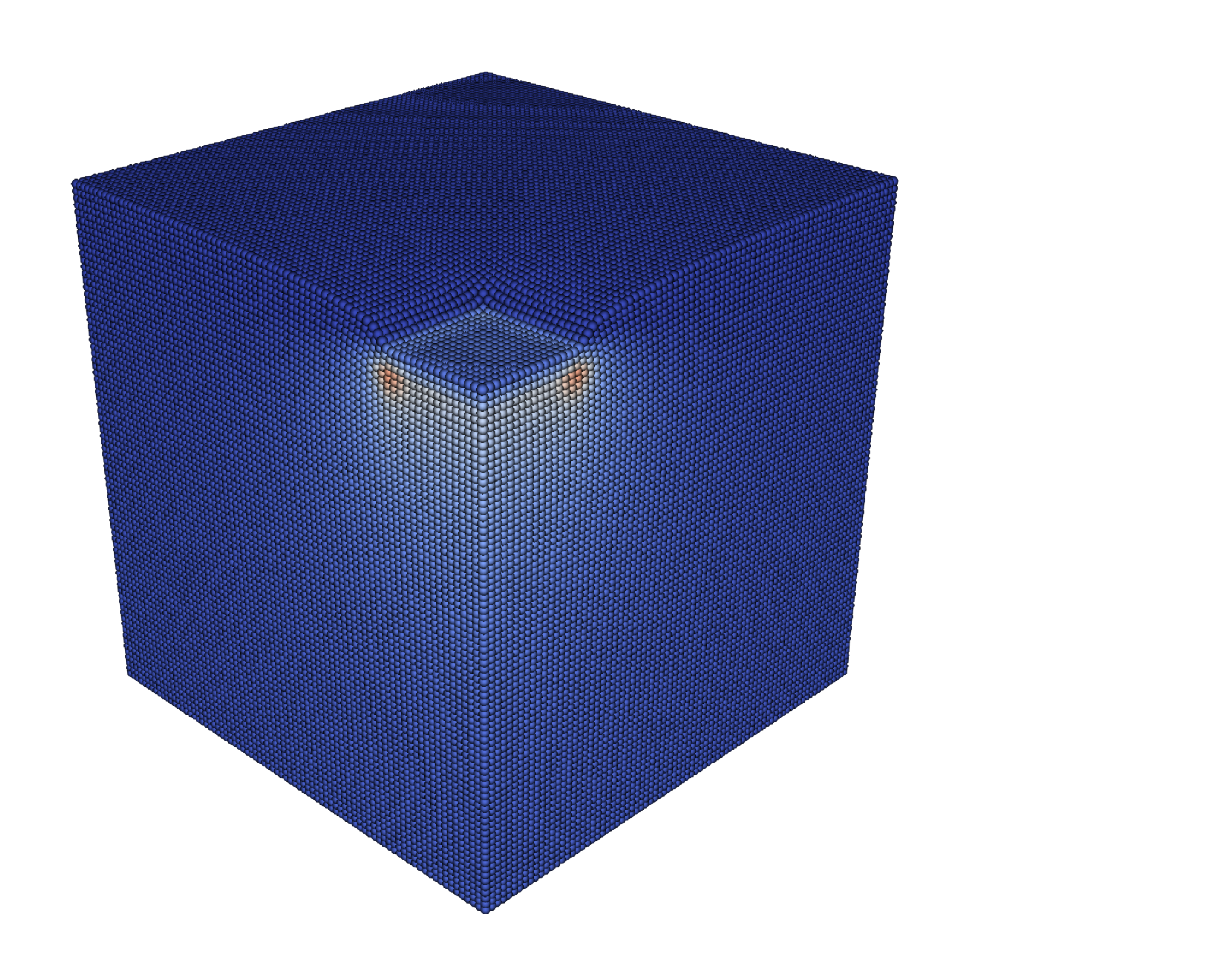}}\\
    \subfloat[Load step 20]{\includegraphics[height=0.43\textwidth]{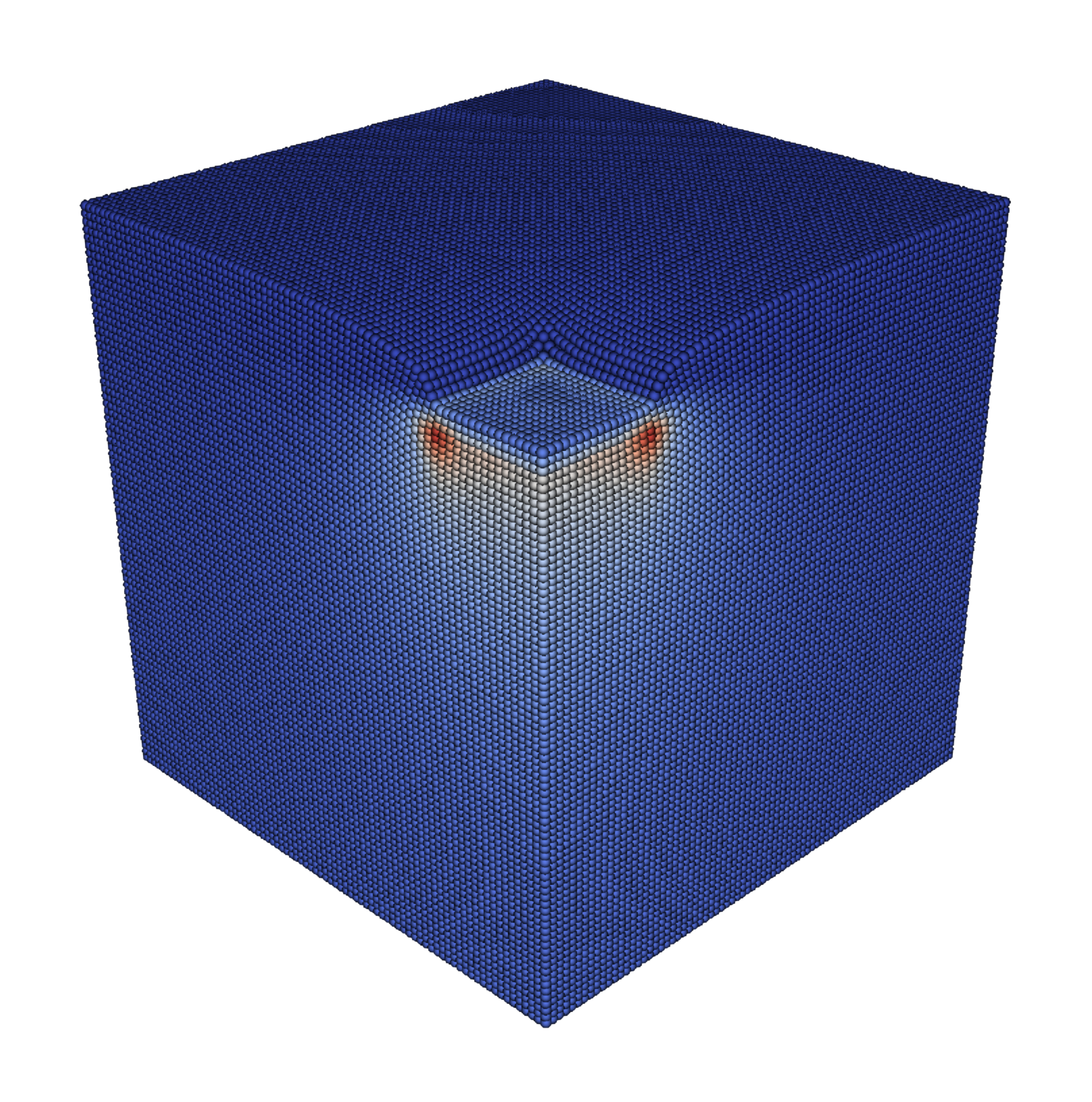}}\hspace{1em}
    \subfloat[Load step 25]{\includegraphics[height=0.43\textwidth]{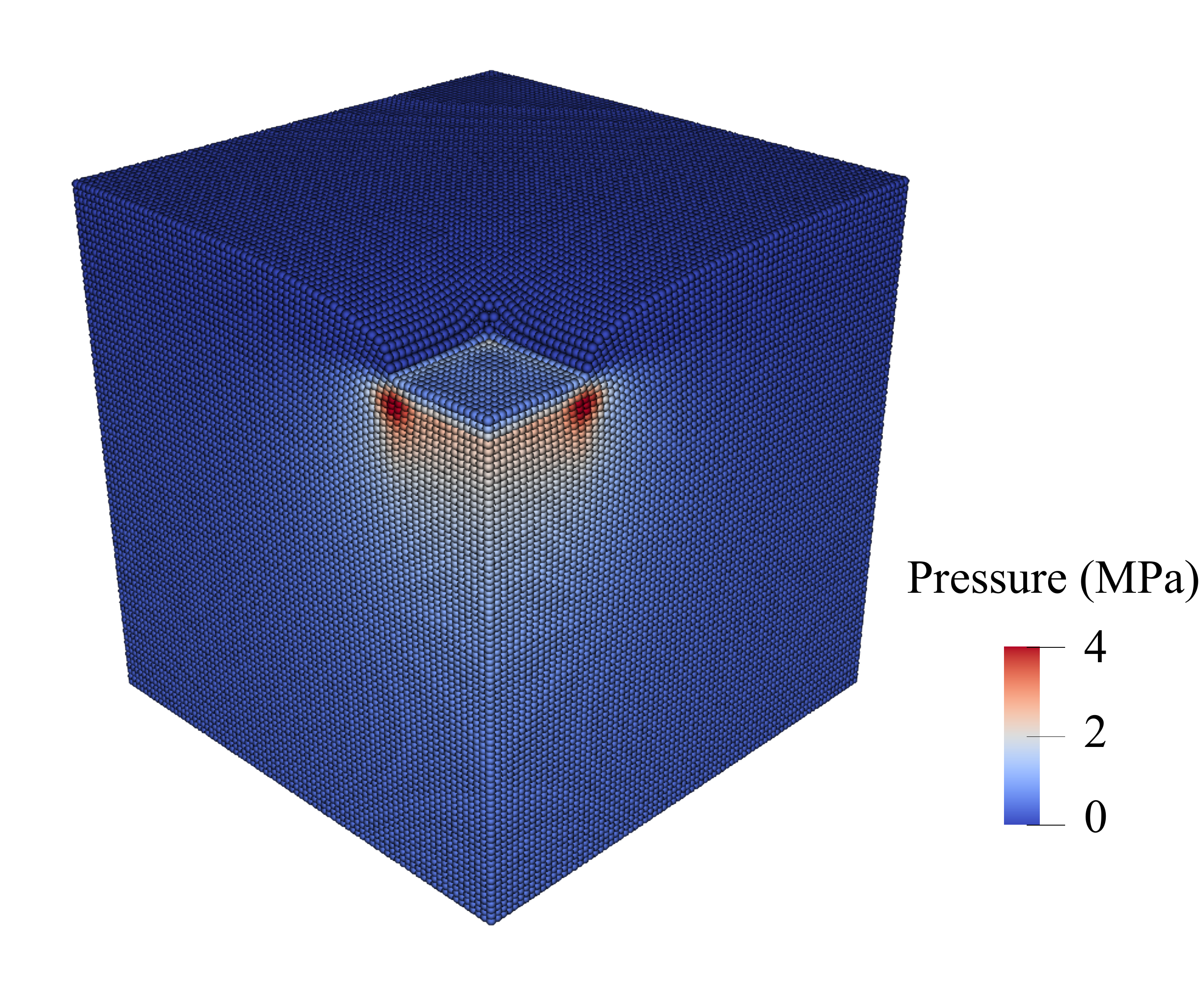}}
    \caption{Ground stiffness identification via inverse analysis: excess pore pressure distributions at selected load steps.}
    \label{fig:indentation_pressure}
\end{figure}

We first examine the system response under forward simulation.
Figure~\ref{fig:indentation_pressure} shows the excess pore pressure distributions at selected load steps as the footing is indented to a final depth of 0.5 m over 25 increments.
As expected, pore pressure builds up beneath the footing due to compression of the saturated ground.
The simulation also captures the expected downward displacement directly beneath the footing and uplift in the surrounding region, manifesting a physically consistent deformation pattern.

\begin{figure}[h!]
  \centering
  \includegraphics[width=0.5\textwidth]{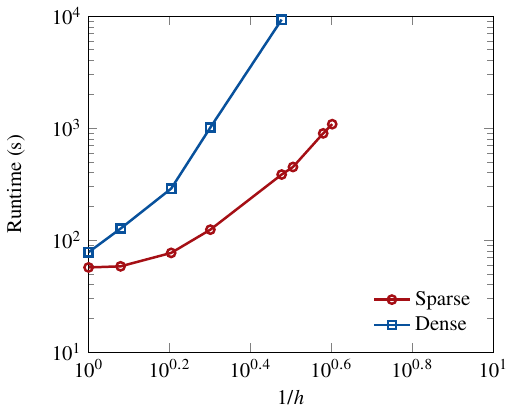}
  \caption{\revised{Ground stiffness identification via inverse analysis: scalability analysis.}}
  \label{fig:indentation_scalability}
\end{figure}

\revised{To further assess the scalability of the proposed framework, we conduct additional simulations at varying discretization levels using both the sparse and dense differentiation algorithms.
As shown in Fig.~\ref{fig:indentation_scalability}, the total runtime increases with mesh refinement for both algorithms; however, the growth rate is substantially lower for the sparse formulation, resulting in a noticeably flatter slope.
This behavior indicates superior scalability of the sparse differentiation with respect to problem size.
The runtime gap between the two formulations becomes more pronounced in this large-scale 3D problem compared to the 2D case presented in Fig.~\ref{fig:cantilever_beam_total_runtime}, where the difference was less significant. This trend highlights the scalability of the sparse formulation and underscores the importance of reducing AD passes---rather than optimizing individual kernel performance---for achieving high-performance Jacobian construction.}

We now turn to the inverse problem.
In this example, the indentation force--displacement response is highly sensitive to the Young's modulus of the ground.
To estimate this parameter, we define a scalar loss function based on the slope of the force--displacement curve.
Specifically, the loss is computed as the squared difference between the slope of the simulated response and that of a reference response generated using the true value $E = 10$ MPa. 
This formulation provides a simple yet effective test case for evaluating the framework’s differentiability and inversion capabilities.
GeoWarp records the full computational graph of the simulation, enabling the gradient of the loss function with respect to Young’s modulus to be computed automatically and exactly via algorithmic differentiation. 
These gradients are then used in a gradient-based optimization algorithm---specifically, gradient descent with a learning rate of 0.2, selected through preliminary tuning.
To test the robustness of the algorithm, the optimization is initialized with a poor initial guess of $E = 1$ MPa---an order of magnitude lower than the true value. 
The optimization proceeds until the loss falls below a prescribed threshold, indicating sufficient agreement between the simulated and reference responses.

\begin{figure}[h!]
    \centering
    \includegraphics[width=0.5\textwidth]{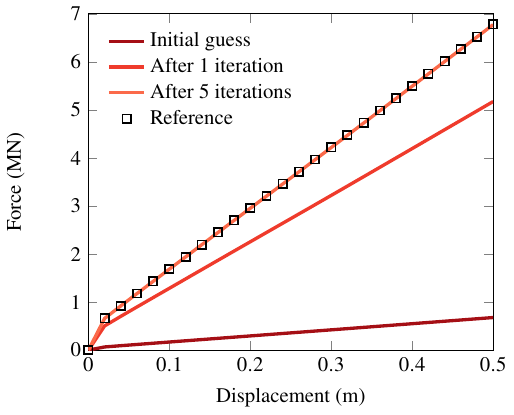}
    \caption{Ground stiffness identification via inverse analysis: force--displacement curves at selected iterations of the inverse analysis, compared with the reference response.}
    \label{fig:indentation_force}
\end{figure}

\begin{figure}[h!]
    \centering
    \includegraphics[width=0.5\textwidth]{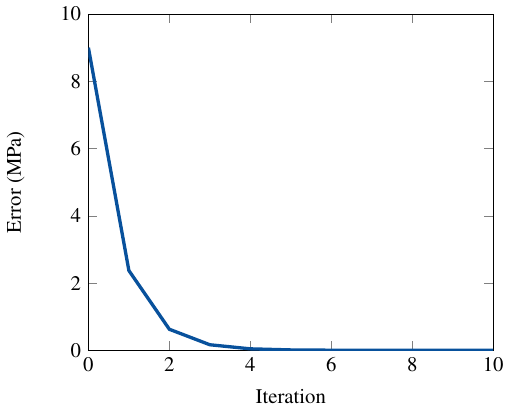}
    \caption{Ground stiffness identification via inverse analysis: evolution of the loss function during inverse analysis.}
    \label{fig:indentation_loss}
\end{figure}

Figure~\ref{fig:indentation_force} shows the simulated force--displacement curves at the initial guess, after one iteration, and after five iterations of gradient descent, compared with the reference response. 
Although the initial guess produces significant discrepancies, the solution rapidly converges toward the reference. 
Figure~\ref{fig:indentation_loss} quantifies this convergence, showing that the loss decreases sharply within just a few iterations.

These results highlight the effectiveness of automatic differentiation in enabling efficient inverse analysis. 
By eliminating the need for manual derivation of sensitivity expressions, GeoWarp simplifies the development of gradient-based parameter identification workflows. 
Although this example involves a single scalar parameter and a simplified loss metric, the same methodology can be generalized to more complex problems involving multiple parameters, spatial heterogeneity, or more sophisticated objective functions.

\section{Closure}
\label{sec:closure}

In this work, we introduced GeoWarp---the first open-source, high-performance, differentiable implicit MPM framework for geomechanics, built on NVIDIA Warp.
GeoWarp integrates AD into an implicit MPM solver, eliminating the need for manual derivation of Jacobian matrices in Newton-type methods.
This capability lowers the barrier to implementing implicit MPM methods, particularly in problems involving complex constitutive models where consistent tangent operators are required.
To address the computational cost of AD, we have developed a Jacobian construction algorithm that leverages the inherent sparsity of the MPM formulation. 
By limiting the number of backward passes to a small, fixed value---independent of problem size---the method enables efficient large-scale simulations on modern GPUs.
We have demonstrated the accuracy and versatility of GeoWarp through forward and inverse examples in large-deformation elastoplasticity and coupled poromechanics.
These results show that combining implicit MPM with AD enables efficient gradient computation and facilitates a range of applications, not only quasi-static or long-term simulations but also constitutive model calibration, parameter identification, and integration with gradient-based optimization workflows. 
GeoWarp thus offers a robust, scalable, and extensible foundation for advancing implicit and differentiable MPM simulations in computational geomechanics.

\revised{Future work will extend the current framework in several directions.
First, we plan to extend it to inverse analysis of problems involving more complex constitutive behavior, where non-differentiable features (\eg~elastic--plastic transitions or nonsmooth yield criteria) give rise to challenges for gradient-based methods.
Second, although all tested cases currently fit within available GPU memory, future developments will explore strategies for handling memory overflow, such as domain decomposition or out-of-core execution.
Finally, a detailed hardware-portability study may be conducted to evaluate performance across different GPU architectures.}

\section*{Author contributions} 
\textbf{Yidong Zhao}: Conceptualization, Methodology, Software, Validation, Formal Analysis, Investigation, Data Curation, Writing - Original Draft, Visualization.
\textbf{Xuan Li}: Methodology, Software, Validation, Writing - Review \& Editing.
\textbf{Chenfanfu Jiang}: Methodology, Validation, Writing - Review \& Editing.
\textbf{Jinhyun Choo}: Conceptualization, Methodology, Validation, Resources, Writing - Original Draft, Writing - Review \& Editing, Supervision, Project Administration, Funding Acquisition.

\section*{Acknowledgments}
This work was supported by the National Research Foundation of Korea (NRF) grant funded by the Korean government (MSIT) (No. RS-2023-00209799).

\section*{Data availability statement}
The data that support the findings of this study are available from the corresponding author upon reasonable request.

\bibliography{references}

\end{document}